\documentstyle[prb,eqsecnum,aps,epsf]{revtex}   

\begin{document}
\draft
\title{Excitons in quasi-one dimensional organics: Strong correlation  		         
approximation}
\author{Z. G. Yu, A. Saxena, and A. R. Bishop}
\address{Theoretical Division, Los Alamos National Laboratory, Los 
Alamos, New Mexico 87545}
\date{\today}
\maketitle
\begin{abstract}
An exciton theory for quasi-one dimensional organic materials is 
developed
in the framework of the Su-Schrieffer-Heeger Hamiltonian augmented by 
short range extended Hubbard interactions. 
Within a strong electron-electron
correlation approximation, the exciton properties are 
extensively studied. Using scattering theory, we analytically obtain
the exciton energy and wavefunction and derive a  
criterion for the existence of a $B_u$ exciton. 
We also
systematically investigate the effect of impurities on the coherent motion 
of an exciton. The coherence is measured by a suitably defined 
electron-hole correlation function. It is shown that, for impurities with
an on-site potential, a crossover behavior will occur if the impurity 
strength is comparable to the bandwidth of the exciton, corresponding to 
exciton localization. For a charged impurity with a spatially extended
potential, in addition to localization the exciton will dissociate into an 
uncorrelated electron-hole pair when the impurity is sufficiently strong  
to overcome the Coulomb interaction which binds the electron-hole pair.
Interchain coupling 
effects are also discussed by considering two polymer chains coupled through
nearest-neighbor interchain hopping $t_{\perp}$ and interchain Coulomb 
interaction 
$V_{\perp}$. Within the $t$ matrix scattering formalism, for every 
center-of-mass
momentum, we find two poles determined only by $V_{\perp}$, which 
correspond to the interchain excitons, and four poles only involving intrachain
Coulomb $V$, which are intrachain excitons. The interchain exciton 
wavefunction is analyzed in terms of inter- and intra-chain character.
Finally, the exciton state is 
used to study the charge transfer from a polymer chain to an adjacent
dopant molecule. 
From a variational wave function for the total system, we explore 
the dependence of the probability of charge transfer on the acceptor level, 
the hopping, and the wavefunction of the exciton.
\end{abstract}

\section{introduction}

In recent years, the exciton concept of electron-hole bound states
has gained popularity in conjugated organic polymers.
Experimentally, it has been discovered that poly(phenylene vinylene) (PPV) and 
its
derivatives, e.g., poly[2-methoxy, 5-(2' ethyl-hexoxy)-1,4 phenylene vinylene]
(MEH-PPV) can be used as the active luminescent layer in electroluminescent 
light-emitting diode devices.\cite{Bu90}
 By using different conjugated polymers, polymer 
light-emitting diodes (PLED) have been fabricated which emit throughout the 
visible region of the spectrum.\cite{Bu90,BH91,Bu92,Be94,Br92,Gu92} 
It is believed that radiative recombination of 
excitons gives rise to luminescence, so a comprehensive understanding of exciton
properties in polymer chains is very important to guide improvements in quantum
efficiency 
of these PLED devices. Also, conjugated polymers have shown potential
application in photonics for their large optical nonlinearity and 
ultrafast response time.\cite{HKSS,CZ87} In view of
the significant role excitons play
in optical properties of a system, a detailed study of the exciton is also an
important issue for polymer photonic device design. Theoretically, excitons in 
conjugated polymers are both attractive and challenging because of the 
coexistence of low-dimensional confinement 
and strong electron-electron ({\em e-e}) correlation in these systems. These
ingredients have in fact led to many 
controversies during the study of excitons in 
polymers.\cite{Ch94,CC93,SBS94,Yu95}

Discussion of excitons in solids can be traced back more than half a century to
the pioneering work by Frenkel\cite{Fr31} which even preceded than the
band theory in solids. After numerous studies on excitons over several decades, 
the exciton, as an elementary excitation, has been well established in bulk
insulators. This is the reason why in the existing exciton 
theories for polymers,\cite{Su84,HN85,TIH87,AYS92}  the
standard exciton theory in semiconductors\cite{NTA80}
was usually borrowed with limited justification. In these theories, the 
polymer is 
regarded as a Peierls insulator which means the gap between the conduction 
and valence bands arises from the Peierls dimerization, a well-known nesting 
effect in one-dimensional metals.\cite{Pe55}
 Then the exciton states are solved considering the {\em e-e}
interaction as a perturbation as in traditional exciton theory. 
But the polymer is significantly different from the conventional semiconductor.
In a semiconductor, the  electron correlation effects can largely be neglected
and it is plausible to treat the
{\em e-e} interaction as a perturbation. However, the polymer
is typically a strongly correlated system with a moderate to large on-site 
Hubbard repulsion, and much of the band gap is due to
the electron correlation rather than the 
dimerization.\cite{OUK72,Uk79,BM85,sun}
Thus the foundation of existing exciton theories in conjugated polymers 
is not so firm, and
these theories have already led to some
qualitatively incorrect results. For example, from these theories,
the 1$B_u$ exciton is lower than the 2$A_g$, but typically
the order should be reversed. Again, the threshold of the conduction
band is independent of $U$, the on-site Hubbard repulsion.\cite{AYS92} 
This is unreasonable 
since, when an electron is excited to the conduction band, double occupation 
must occur and cause an additional Hubbard repulsion.

Having appreciated the central importance of electron correlation in polymers,
several efforts have been undertaken to take account of the strong correlation 
effect on exciton states.\cite{Shuai,Gu93,SEGR,Ma96,GCM95}  
In these works, numerical exact diagonalization and density matrix
renormalization group (DMRG) approaches are employed to handle small systems. 
From those numerical
results, some useful information on the electron structure of the polymer can 
be captured
by extrapolating the results to a long chain. However, these works cannot 
be regarded as constituting an exciton theory, since only some specific states 
are focused on in the
calculations and the exciton is not treated as a quasi-particle. We must seek 
a more complete 
exciton theory of conjugated polymers, in which the correlation is stressed 
and the exciton is an elementary 
excitation. This kind of theory will then be useful to study the optical and 
transport properties related to the exciton. In this paper, we will develop a 
new exciton theory 
in the limit in which the Hubbard $U$ is the main origin of the band gap (i.e.,
the polymer is regarded as a Mott insulator). In this regime, the spin has 
little effect on the exciton states and the energy difference between the 
singlet and triplet states is negligibly small compared with their binding 
energies. Using scattering theory, we will analytically calculate the exciton 
states and find
a critical strength of the {\em e-e} interaction for the existence of bound 
exciton states.

Currently, chemically synthesized polymers cannot be free from impurities and 
the ``pristine'' samples
of polymers contain a non-negligible density of impurities and defects from
the cross-linking, complex morphological 
effects, conjugation length effects, and some extrinsic defects. Such 
impurities sometimes critically
influence the properties of the system, e.g., the transport. For 
example, the conductivity in {\em trans}-polyacetylene can be dramatically 
enhanced by 13 orders of magnitude by doping.\cite{HKSS}
In an impurity-free system, the exciton states with different 
center-of-mass momenta form exciton bands, and within the band the exciton 
moves coherently as a composite particle. The disorder tends to produce 
localized 
states, and in one-dimensional systems, any nonvanishing impurity potential
will lead to a localized electronic state.\cite{AALR} 
Thus it is interesting to examine
the localization of an exciton, a composite particle, due to the impurities, 
and furthermore to determine
if the exciton ceases to be a composite particle of electron and hole when the 
impurity is strong enough. In this paper,
we will address the interplay of coherent motion of the exciton and different
types of impurities in our conjugated polymer model.

Strictly speaking, the polymer is only a quasi-one dimensional system, in which
interchain couplings always exist,
and sometimes their effects are striking. Many calculations have
shown that nonlinear excitations like solitons and polarons may be unstable
by taking the interchain coupling into account.\cite{BM83,MC93}
 Recently, great attention has 
been paid to the interchain effects in luminescent polymers, since many 
experiments demonstrated that a large
fraction of primary photoexcitations are interchain 
excitons or polaron pairs.\cite{HYJR,Yan,YRKM} 
Also some theoretical calculations have been carried out to explore the 
interchain coupling
effects on exciton properties.\cite{MC94,Yu96}
Although the terminology of interchain exciton and
intrachain exciton are widely used in current literature, these concepts are 
not so clearly delineated.
From the principles of quantum mechanics, the wavefunction of every eigenstate
in the coupled 
system must be distributed over the whole system, so it is difficult to 
distinguish from the 
wavefunction which one corresponds to the interchain and which one to the 
intrachain exciton. We will clarify what the interchain and intrachain exciton 
states are, and calculate their energies and wavefunctions.

Photoconductivity in conjugated polymers can sometimes be greatly enhanced by 
intercalating or doping the polymer with a particular species of molecule.
Interesting examples of this phenomenon occur when MEH-PPV is doped by 
fullerene C$_{60}$ molecules.\cite{Sa92,KC93}
This is because the exciton, the bound electron and hole state in the polymer,
will decay when the dopant molecule is introduced. The electron (or hole) in 
the exciton will transfer from the polymer chain to the doped molecule,
giving rise to a free carrier. Rice and Gartstein recently
proposed a theory to explain the ultrafast time scale for this charge 
transfer.\cite{RG96} From a 
quantum mechanics perspective, assuming we have an exciton state in the 
polymer chain due to
photoexcitation, when the coupling between the chain and the molecule is 
switched on, the electron will move in the whole system (including the chain 
and the 
molecule), and this state must have a lower energy than the initial state. 
So another
point of view from which to study the charge transfer is to ask
what percentage of the electron (hole) has transferred from
the polymer chain to the dopant. This percentage should depend on the acceptor 
level and the coupling between the chain and the molecule, as well as the 
initial exciton wavefunction in the polymer chain. We will discuss this issue 
here.

The paper is organized as follows. First, we develop an exciton theory 
for conjugated polymers in the
strong correlation (large Hubbard $U$) approximation in Sec. II.
In Sec. II.A we simplify a Peierls-extended
Hubbard model to a model represented by spinless fermions with short-range
{\em e-e} interactions in real space. Then we use $t$ matrix scattering theory
to determine 
the wavefunction and binding energy of exciton states analytically and derive
a criterion for the existence of the $B_u$ exciton in Sec. II.B. This criterion
is further proved according to the Levinson's theorem in scattering theory  
in Sec. II.C. In Sec. II.D a more formal and compact formalism for optical 
absorption 
in conjugated polymers is presented based on our exciton theory in the
large-$U$ limit. Sec. III is devoted
to the impurity effects on the coherent motion of the exciton. Using a suitably
defined electron-hole correlation function, we study different types of 
impurity. In Sec. IV,
we investigate interchain coupling effects by considering a two-chain system
supplemented by nearest-neighbor interchain hopping $t_{\perp}$ and 
interchain {\em e-e} interaction $V_{\perp}$. Using $t$ matrix formalism, we 
analytically determine the poles corresponding to intrachain and interchain
excitons, respectively. We also show the wavefunction of the interchain 
exciton.
In Sec. V, the static $A_g$ and $B_u$ excitons are used to study the charge 
transfer 
in a  molecularly-doped polymer. By constructing a variational wavefunction
for the whole system, the energy of this variational state, and accordingly the
probability of charge transfer, can be obtained. Finally, we summarize
our results in Sec. VI.

\section{A new exciton theory in conjugated polymers}

In existing theories, the polymer is regarded as a Peierls insulator, and then 
the exciton state is determined by 
treating the {\em e-e} interaction (including the on-site Hubbard 
interaction) as a perturbation. In this picture, the single particles 
(electron and 
hole) are defined based on a non-interacting Su-Schrieffer-Heeger model,
so the band gap, from these theories,
is independent of the {\em e-e} interaction. In a strongly correlated
system, the electronic states are quite different from the non-interacting 
model. Since
the ground state is half-filled, an electron excited to the conduction band 
must
cost the additional Hubbard repulsion energy caused by the double occupation. 
In conjugated polymers and related organic conductors, it is now 
accepted that the origin of the band gap comes typically from the Hubbard 
repulsion rather
than the Peierls dimerization. So the Hubbard term should be given priority
when one develops an exciton theory. In this section, we will 
regard the polymer as a Mott insulator and develop the exciton theory in 
large-$U$ limit. In this energy regime, double and higher order electron-hole
excitations can be neglected because of their high energies ($\ge 2U$), and a
single configuration interaction approximation is reasonable in determining 
exciton states. Before carrying out the calculation,
let us recall how large the Hubbard $U$ (in units of electronic hopping energy
$t$) is in real systems: $U/t\sim 3-4$ in conjugated polymers
and $U/t\sim 8-10$ in segregated stack charge transfer salts.\cite{BCM92,Soos}
Strictly, this strong correlation limit is applicable only when $U\gg t$, 
thus real conjugated polymers only marginally satisfy this approximation.

\subsection{Hamiltonians}

The theoretical model we consider is the Peierls-extended Hubbard model, i.e.,
the Su-Schrieffer-Heeger\cite{HKSS}
model augmented by an extended Hubbard interaction. For a one-dimensional 
chain, this model Hamiltonian is
\begin{equation}
H=-t\sum_{l\sigma}[1-(-
1)^l\delta](c^{\dagger}_{l\sigma}c_{l+1\sigma}+{\rm
H.c.})+U\sum_l
n_{l\uparrow}n_{l\downarrow}+V\sum_l(\rho_l-1)(\rho_{l+1}-1)~.
\end{equation}
Here $c^{\dagger}_{l\sigma}$ creates an electron of spin $\sigma$ on 
site
$l$, $t$ is the one-electron hopping integral, $\delta$ is a
bond-alternation parameter, $U$ and $V$ are respectively
the on-site and nearest-neighbor
Coulomb interaction, $n_{l\sigma}=c^{\dagger}_{l\sigma}c_{l\sigma}$ is the
number operator, and
$\rho_l=n_{l\uparrow}+n_{l\downarrow}$. Since we will be concerned only with
electronic excitations in this work, we consider a rigid dimerized ground state
as a starting point, and do not specify its explicit origin (e.g., {\em e-e}
interactions, electron-phonon couplings, or crystal structure).
Strictly speaking, this Peierls-extended Hubbard model is directly applicable 
only to 
{\em trans}-polyacetylene.
However, recent calculations have shown that the primary excitation in
luminescent polymers like PPV can also be described within an effective linear
chain model.\cite{SEGR} In these luminescent polymers, the lowest
excitonic wave function extends over several repeat units.
The properties of exciton are therefore not very sensitive to the 
detailed structure within the unit cell. From the viewpoint of renormalization,
we can map the complex structure of a luminescent polymer into an effective 
Peierls-extended Hubbard system with the
same significant physical properties by integrating out the superfluous degrees
of freedom
caused by the complicated unit cell structure. 
 We have also neglected lattice relaxation,
since many experiments and theories have demonstrated that 
{\em e-e} interactions dominate electron-lattice interactions 
in many luminescent polymers.\cite{Le94,Co90,GB91,RBBH}
This simplification enables us to handle {\em e-e}
interactions in long
chains and arrive at an understanding of {\em electronic states} in
conjugated polymers without loss of essential physics, although the
quantitative explanation of some {\em lattice} property like vibronic
structure or bond length should, indeed, take into account
lattice relaxation effects.\cite{BSFB}

To emphasize the electron correlation, we begin with the Hubbard model
\begin{equation}
H_0=-t\sum_{l\sigma}(c^{\dagger}_{l\sigma}c_{l+1\sigma}+{\rm
H.c.})+U\sum_l n_{l\uparrow}n_{l\downarrow}~.
\end{equation}
Although the exact wave function and the ground-state energy of this
Hamiltonian have been obtained by Lieb and Wu,\cite{LW68} the 
Green's function and
correlation functions are difficult to calculate directly by using the 
exact
wave function, and it is also difficult to use their solution to study
the exciton. As a practical alternative, here we
make the strong correlation (large-$U$) approximation. In this approximation, 
as we will see later, the band gap is essentially $U$, which is not the 
same as the exact solution for the Hubbard model by Lieb and Wu,\cite{LW68}
where the charge excitation gap is $U-4t+\sum_{n=1}^{\infty}(-1)^n[\frac{1}{2}
nU-(t^2+\frac{1}{4}n^2U^2)^{1/2}]$. However, for the strong correlation 
limit $U\gg 4t$, this difference is not important, and does not affect the exciton 
trends we wish to establish.

The density product $n_{l\uparrow}n_{l\downarrow}$ can be expressed by
the on-site electron number and spin operator, and the Hubbard model is 
rewritten as
\begin{equation}
H_0=-t\sum_{l\sigma}(c^{\dagger}_{l\sigma}c_{l+1\sigma}+{\rm
H.c.})+U\sum_l [\frac{1}{2}\rho_l-(-1)^l{\rm\bf S}_l\cdot{\rm\bf 
n}_l]~,
\end{equation}
where ${\bf S}_l$ is the electron spin operator at site $l$,
\begin{equation}
{\bf S}_l=\frac{1}{2}\sum_{\sigma \sigma'}c^{\dag}_{l\sigma}
{\boldmath\sigma}_{\sigma \sigma'}c_{l\sigma'}~,
\end{equation}
and ${\bf n}_l$ is a unit vector along
the spin polarization axis of the electron.\cite{WSTS} 

When $U\gg t$, the ground state is expected to have N\'{e}el 
order. We make an
approximation by assuming that {\bf n}$_l$ always coincides with the 
$z$ axis, which implies that the spin excitations have been ignored. Thus, in 
this approximation,
the singlet and triplet excitons have the same energy and are not
distinguishable.
This is 
reasonable since  for a one-dimensional Hubbard model, the spin
and charge excitations are separated when $U\to \infty$ and by order $U/t$ for
$U\gg t$, and the exciton is a charge excitation. This point is also directly
justified 
by comparing energies of the singlet and triplet states obtained from a finite 
chain exact diagonalization calculation.\cite{note}

\begin{equation}
H_0\simeq
-t\sum_{l\sigma}(c^{\dagger}_{l\sigma}c_{l+1\sigma}+{\rm
H.c.})+\frac{U}{2}\sum_{l\sigma}[1-(-
1)^l\sigma]c^{\dagger}_{l\sigma}c_{l\sigma}~.
\end{equation}
This Hamiltonian is readily diagonalized by introducing
\begin{equation}
c_{l\sigma}=\frac{1}{\sqrt{N}}{\sum_k}' e^{ikl}[(u_k+\sigma
p_lv_k)\alpha_{k\sigma}+p_l(u_k-\sigma p_lv_k)\beta_k]~.
\end{equation}
Here the prime means that the summation runs over the reduced 
Brillouin
zone $|k|<\pi/2$ and $p_l=(-1)^l$. Then
\begin{equation}
H_0=-{\sum_{k\sigma}}'\Bigl[(E_k-
\frac{U}{2})\alpha^{\dagger}_{k\sigma}
\alpha_{k\sigma}+(E_k+\frac{U}{2})\beta^{\dagger}_{k\sigma}
\beta_{k\sigma}\Bigr]~,
\end{equation}
with
\begin{equation}
E_k=\sqrt{\frac{U^2}{4}+\varepsilon^2_k}~,
\end{equation}
\begin{equation}
\varepsilon_k=-2t\cos k~.
\end{equation}
The functions $u_k$ and $v_k$ are 
\begin{equation}
u_k=\frac{1}{\sqrt{2}}\sqrt{1+\frac{|\varepsilon_k|}{E_k}}
\simeq\frac{1}{\sqrt{2}}\Bigl(1+\frac{|\varepsilon_k|}{U}\Bigr)~,
\end{equation}
\begin{equation}
v_k=\frac{1}{\sqrt{2}}\sqrt{1-\frac{|\varepsilon_k|}{E_k}}
\simeq\frac{1}{\sqrt{2}}\Bigl(1-\frac{|\varepsilon_k|}{U}\Bigr)~.
\end{equation}

In the case of $U\gg t$, a localized picture is more 
convenient. Two spinless fermions can be defined in the lattice representation
as follows\cite{WSTS}
\begin{eqnarray}
\alpha_l&=&\sum_{\sigma}\theta(p_l\sigma)\sqrt{2/N}{\sum_k}'
e^{ikl}\alpha_{k\sigma}~,\\
\beta_l&=&\sum_{\sigma}\theta(p_l\sigma)\sqrt{2/N}{\sum_k}'
e^{ikl}\beta_{k\sigma}~,
\end{eqnarray}
where $\theta(x)$ is the step function. Expanded in
powers of $t/U$, $c_{l\sigma}$ can be expressed by $\alpha_l$ and
$\beta_l$:
\begin{equation}
c_{l\sigma}=\theta(p_l\sigma)\alpha_l+\theta(-p_l\sigma)p_l\beta_l
+\theta(-p_l\sigma)\frac{t}{U}(\alpha_{l+1}+\alpha_{l-1})
+\theta(p_l\sigma)p_l\frac{t}{U}(\beta_{l+1}+\beta_{l-
1})+O(\frac{t^2}{U^2})~.
\end{equation}

If we include the bond alternation part in our unperturbed Hamiltonian, then
\begin{eqnarray}
H'_0&=&H_0+\sum_{l\sigma}(-1)^l\delta
t(c^{\dagger}_{l\sigma}c_{l+1\sigma}+{\rm H.c.})\nonumber\\
&=&J\sum_l(h^{\dagger}_lh_l+\beta^{\dagger}_l\beta_l)+
U\sum_l\beta^{\dagger}_l\beta_l+\frac{J}{2}\sum_l(h^{\dagger}_{l+2}h_l
+\beta^{\dagger}_l\beta_{l+2}+{\rm H.c.})\nonumber\\
&+&\delta t\sum_l(-h_l\beta_{l+1}+h_{l+1}\beta_l
-\beta^{\dagger}_{l+1}h^{\dagger}_l+\beta^{\dagger}_lh^{\dagger}_{l+1})~.
\end{eqnarray}
Here we have introduced the hole operator $h^{\dagger}_i=\alpha_i$, and 
$J=2t^2/U$.

By introducing the Fourier transformations
\begin{eqnarray}
h_l&=&\frac{1}{\sqrt{N}}\sum_k e^{ikl}h_k~,\\
\beta_l&=&\frac{1}{\sqrt{N}}\sum_k e^{-ikl}\beta_{-k}~,
\end{eqnarray}
we can rewrite $H'_0$ in momentum space as
\begin{equation}
H'_0=\sum_k[(J+J\cos 2k)h^{\dagger}_kh_k+(U+J+J\cos
2k)\beta^{\dagger}_k\beta_k+2i\delta t\sin
k(h_k\beta_{-k}-\beta^{\dagger}_{-k}h^{\dagger}_k)]~.
\label{h0'}
\end{equation}

Making the Bogoliubov transformation
\begin{eqnarray}
\eta_k&=&\cos\theta_kh_k+i\sin\theta_k\beta^{\dagger}_{-k}~,\\
\gamma^{\dagger}_{-k}&=&-
i\sin\theta_kh_k+\cos\theta_k\beta^{\dagger}_{-k}~,
\end{eqnarray}
Hamiltonian (\ref{h0'}) can be diagonalized if the relation
\begin{equation}
\sin2\theta_k=\frac{-4\delta t\sin k}{U}
\end{equation}
is satisfied, yielding
\begin{equation}
H'_0=\sum_k(\epsilon_k\eta^{\dagger}_k\eta_k+\tilde{\epsilon}_k
\gamma^{\dagger}_k\gamma_k)~,
\end{equation}
with
\begin{eqnarray}
\epsilon_k&=&J(1+\delta^2)+J(1-\delta^2)\cos 2k~,\\
\tilde{\epsilon}_k&=&U+J(1-3\delta^2)+J(1+3\delta^2)\cos 2k~.
\end{eqnarray}
Operators $\eta^{\dagger}_k$ and $\gamma^{\dagger}_k$ create the hole and 
electron in the new valence and conduction band, respectively.
Their lattice representations are
\begin{eqnarray}
\eta_l&=&\frac{1}{\sqrt{N}}\sum_k e^{ikl}\eta_k~,\\
\gamma_l&=&\frac{1}{\sqrt{N}}\sum_k e^{-ikl}\gamma_{-k}~,
\end{eqnarray}
which can be expressed by $h_l$ and $\beta_l$ to order $1/U$:
\begin{eqnarray}
\eta_l\simeq h_l-\frac{\delta
t}{U}(\beta^{\dagger}_{l+1}-\beta^{\dagger}_{l-1})~,\\
\gamma^{\dagger}_l\simeq \beta^{\dagger}_l+\frac{\delta
t}{U}(h_{l+1}-h_{l-1})~.
\end{eqnarray}

The inter-site Coulomb interaction is necessary to bind the electron 
and
hole. We consider the $V$-term in the Peierls-extended Hubbard model
as a scattering potential, which has the local representation
\begin{eqnarray}
H_{\rm int}&=&V\sum_{l}(\rho_l-1)(\rho_{l+1}-1)\nonumber\\
&=&V\sum_l(h^{\dagger}_{l+1}h^{\dagger}_lh_lh_{l+1}
+\beta^{\dagger}_{l+1}\beta^{\dagger}_{l}\beta_l\beta_{l+1}
-h^{\dagger}_{l+1}\beta^{\dagger}_l\beta_lh_{l+1}
-\beta^{\dagger}_{l+1}h^{\dagger}_lh_l\beta_{l+1})~.
\end{eqnarray}

Since the main interest here is an exciton, only the interaction between the 
electron and hole is relevant. To order $1/U$, we have
\begin{equation}
H^{\rm e-h}_{\rm int}=-V\sum_l(\eta^{\dag}_{l+1}\gamma^{\dag}_l\gamma_l
\eta_{l+1}
+\gamma^{\dag}_{l+1}\eta^{\dag}_l\eta_l\gamma_{l+1})~.
\end{equation}

\subsection{Exciton states: $t$ matrix theory}

Since the Hamiltonian is invariant with respect to translation, the
exciton states can be classified according to the total quasimomentum
$K$. We can write the exciton wave function as
\begin{equation}
|\Psi_K \rangle=\sum_sB_{s,K}|\psi_{s,K}\rangle~,
\end{equation}
where $K$ is the center-of-mass momentum. The basis is chosen as
\begin{equation}
|\psi_{s,K}\rangle=\frac{1}{\sqrt{N}}\sum_le^{iKl}\gamma^{\dagger}_{l+s}
\eta^{\dagger}_l|g\rangle~,
\label{bas}
\end{equation}
representing a created electron-hole pair from the ground state
$|g\rangle$ with a separation $s$ in real space.
We will determine the exciton state by using $t$ matrix scattering theory.
According to $t$ matrix theory\cite{Ca91}
\begin{equation}
{\cal T}(z)={\cal U}+{\cal UG}(z){\cal T}(z)~,
\label{t=}
\end{equation}
where ${\cal T}(z)$ is the $t$ matrix, ${\cal G}(z)$ the notation for
resolvent $1/(z-H'_0)$, and ${\cal U}$ the potential operator.
Equation (\ref{t=}) has the formal solution
\begin{equation}
{\cal T}(z)={\cal U}/[1-{\cal G}(z){\cal U}]~.
\label{t=1}
\end{equation}
Using the
basis of Eq. (\ref{bas}), we obtain the Green's function
\begin{eqnarray}
(r|{\cal G}(z)|s)&\equiv&G(r-s;z)\nonumber\\
&=&\langle\psi_{r,K}|(z-H'_0)^{-1}|\psi_{s,K}\rangle
=\frac{1}{N}\sum_k
\frac{e^{ik(r-s)}}{z-(\tilde{\epsilon}_k+\epsilon_{-k+K})}~.
\label{gre}
\end{eqnarray}
Here $z=E_K+i0^+$ and the potential matrix is
\begin{eqnarray}
(s|{\cal U}|s')&\equiv&\langle \psi_{s,K}|H^{\rm e-h}_{\rm 
int}|\psi_{s',K}\rangle\nonumber\\
&=&-V\delta_{ss'}(\delta_{s,-1}+\delta_{s,1})~.
\end{eqnarray}

The utility of Eq. (\ref{t=1}) rests on the possibility of actually
constructing the inverse operator $1/(1-{\cal GU})$. This can
be achieved exactly in our case since, conveniently, the potential
is of short range in the local representation. Actually,
the portion of the potential ${\cal U}$ 
containing nonzero elements forms a $2\times 2$ submatrix
under the basis Eq. (\ref{bas}),

\begin{equation}
\left( \begin{array}{cc}
{\cal U}_{-1-1}  & {\cal U}_{-11}\\
{\cal U}_{1-1} & {\cal U}_{11}
\end{array} \right)
=-V\left( \begin{array}{cc}
1  & 0\\
0 & 1
\end{array} \right)~.
\end{equation}
Denoting $D$ as the
determinant of $1-{\cal GU}$, we have
\begin{equation}
D(E_K)=\left| \begin{array}{cc}
1+G(0;E_K)V  & G(-2;E_K)V\\
G(2;E_K)V & 1+G(0;E_K)V
\end{array} \right|~.
\end{equation}
The determinant will vanish for some specific values of the energy, 
which are the energies of the localized states. Consequently, to find the 
energy $E_K$ of the bound exciton state we look for the root of
\begin{equation}
D(E_K)=0~.
\end{equation}
Subsequently, the wave function is calculated by solving the equation
\begin{equation}
B_{r,K}=\sum_{st}(r|G(E_K)|s)(s|{\cal U}|t)B_{t,K}~.
\end{equation}

First, let us focus on the static exciton, i.e., $K=0$. In this case, the
system is symmetric with respect to spatial inversion. Introducing
the transformation
\begin{eqnarray}
B^+_l&=&\frac{1}{\sqrt{2}}(B_l+B_{-l})~,\\
B^-_l&=&\frac{1}{\sqrt{2}}(B_l-B_{-l})~,
\end{eqnarray}
where $B_l \equiv B_{l,K=0}$, and noticing
\begin{equation}
G(s-t;E_0)=G(t-s;E_0)~,
\end{equation}
we can write the determinant $D$ as the product of two parts:
\begin{equation}
D(E_0)=D_-(E_0)D_+(E_0)~.
\end{equation}
Here
\begin{equation}
D_-(E_0)=1+[G(0;E_0)-G(2;E_0)]V
\end{equation}
is for the $A_g$ state with the wave function
\begin{equation}
B^-_l=-V[G(l-1;E_0)-G(l+1;E_0)]B^-_1~;
\end{equation}
and
\begin{equation}
D_+(E_0)=1+[G(0;E_0)+G(2;E_0)]V
\end{equation}
is for the $B_u$ state with the wave function
\begin{equation}
B^+_l=-V[G(l-1;E_0)+G(l+1;E_0)]B^+_1~.
\end{equation}

We denote $x$ as the exciton binding energy:
\begin{equation}
x=E_G-E_0~,
\end{equation}
where $E_G$ is the band gap:
\begin{equation}
E_G=(\tilde{\epsilon}_k+\epsilon_k)|_{k=\pi/2}=U-4\delta^2J~.
\label{gap}
\end{equation}
When $\delta=0$, $E_G$ does not equal the exact result by Lieb and 
Wu \cite{LW68}. However, this is not a problem for estimating correct 
binding energies of the exciton states, since we will directly calculate 
the binding energy rather than first calculating the energy of the exciton.
The Green's functions are readily calculated; for $x>0$
\begin{equation}
G(2l+1;E_0)\equiv 0~,
\end{equation}
\begin{equation}
G(2l;E_0)=\frac{(-1)^{l+1}}{\sqrt{x^2+4J(1+\delta^2)x}}
\Bigl(\frac{1-\sqrt{1-u^2}}{u}\Bigr)^{2l}~,
\end{equation}
with
\begin{equation}
u^2=\frac{4J(1+\delta^2)}{x+4J(1+\delta^2)}<1~.
\end{equation}

Then, by requiring $D_-(E_0)=0$, we obtain the binding energy of
the $A_g$ exciton as
\begin{equation}
x_-=\frac{V^2}{V+J(1+\delta^2)}~,
\end{equation}
with corresponding wave function
\begin{eqnarray}
B^-_{2l-1}&=&-\frac{G(2l;E_0)-G(2l-2;E_0)}{\sqrt{-2[G'(0;E_0)-
G'(2;E_0)]}}~,\\
B^-_{2l}&=&0~.
\end{eqnarray}
Here $G'(n;E)\equiv \frac{d}{dE}G(n;E)$.

For the $B_u$ state,  when $V>2J(1+\delta^2)$,
the binding energy of the exciton is
\begin{equation}
x_+=\frac{[V-2J(1+\delta^2)]^2}{V-J(1+\delta^2)}~,
\end{equation}
and the wave function is 
\begin{eqnarray}
B^+_{2l-1}&=&-\frac{G(2l;E_0)+G(2l-2;E_0)}{\sqrt{-
2[G'(0;E_0)+G'(2;E_0)]}}~,\\
B^+_{2l}&=&0~.
\end{eqnarray}
When $V<2J(1+\delta^2)$,  a
solution $E_0<E_G$ satisfying equation $D_+(E_0)=0$
cannot be found, i.e.,
there is no bound state. This result gives a criterion
for the existence of a $B_u$ exciton. Finite chain numerical DMRG and exact 
diagonalization
calculations show that the binding energy of the 1$B_u$ state is near
zero when $V$ is not large, but are not conclusive as to whether the state is
strictly bound or free.\cite{shuai2} Our analytical results clearly indicate 
that the stable
1$B_u$ exciton does not exist when $V$ is less than $2J(1+\delta^2)$. Since 
this critical value is half of the width of the continuum band, 
$4J(1+\delta^2)$, and the
bandwidth describes the kinetic energy of a free particle,
this criterion is a reflection of the competition between the kinetic energy 
and the attraction of the electron and hole.

\subsection{Proof of the criterion: Levinson's theorem}

The criterion derived above can be proven by Levinson's theorem.\cite{Ne66}
Namely, the number of bound states in representation $s$ 
lying either above or below the continuum band can be determined 
by using
\begin{equation}
\delta_s(E_i)-\delta_s(E_f)=\pi n_s~.
\label{lev}
\end{equation}
Here $E_i$ is the lowest energy in the band and $E_f$ the highest.
$n_s$ is number of states in any row of representation $s$ separated
from the band. $\delta_s$ is the phase shift that appears in the 
usual partial wave expansion for the scattering amplitude,\cite{Ca91}
which can also be extracted from the 
subdeterminants of $\det(1-{\cal GU})$,
\begin{equation}
\tan \delta_s=-\Im D_s/\Re D_s~.
\end{equation}
In the exciton case,
\begin{eqnarray}
E_i&=&E_G~,\\
E_f&=&(\epsilon_k+\tilde{\epsilon}_k)_{k=0}=E_G+4J(1+\delta^2)~.
\end{eqnarray}
It should be noted that in a one-dimensional system the density of states at the 
edge of the continuum band  will diverge, so the form of Eq. (\ref{lev}) in our
case is
\begin{equation}
\delta_s(E_i-0^+)-\delta_s(E_f+0^+)=\pi n_s~.
\end{equation}

The Green's functions in different energy regions can be calculated according 
to the 
definition Eq. (\ref{gre}). We give the explicit expressions in Appendix A. 
Then the 
subdeterminants $D_-$ and $D_+$ are obtained by straightforward calculations.

For $E_0=E_i-0^+$, i.e., the energy is just below the onset of 
the continuum band,
\begin{eqnarray}
D_-&=&-\infty-i0^+~,\\
D_+&=&1-\frac{V}{2J(1+\delta^2)}-i0^+~,
\end{eqnarray}
and the phase shifts are
\begin{equation}
\delta_-(E_i-0^+)=\pi~,
\end{equation}
\begin{equation}
\delta_+(E_i-0^+)=\left\{\begin{array}{lc}
0~~~V<2J(1+\delta^2)\\
\pi~~~V>2J(1+\delta^2)~.
\end{array}\right.
\end{equation}

For $E_0=E_f+0^+$, i.e., the energy is just above the top of the continuum 
band,
\begin{eqnarray}
D_-&=&1+\frac{V}{2J(1+\delta^2)}-i0^+~,\\
D_+&=&+\infty-i0^+~,
\end{eqnarray}
and the corresponding phase shifts are
\begin{equation}
\delta_-(E_f+0^+)=0~,
\end{equation}
\begin{equation}
\delta_+(E_f+0^+)=0~.
\end{equation}

Thus we have
\begin{equation}
\delta_-(E_i-0^+)-\delta_-(E_f+0^+)=\pi~,
\end{equation}
showing that there is always an $A_g$  bound exciton for a nonvanishing 
$V$. However, when $V<2J(1+\delta^2)$,
\begin{equation}
\delta_+(E_i-0^+)-\delta_+(E_f+0^+)=0~,
\end{equation}
indicating that there is no $B_u$ bound exciton. The
bound exciton will appear only when $V>2J(1+\delta^2)$, since then
\begin{equation}
\delta_+(E_i-0^+)-\delta_+(E_f+0^+)=\pi~.
\end{equation}

To get an overall picture of the phase shift in this one-dimensional system, 
we also examine the phase shifts at $E_i+0^+$ and $E_f-0^+$.
For $E_0=E_i+0^+$, i.e., the energy is just above the onset of the continuum 
band,
\begin{eqnarray}
D_-&=&1+\frac{V}{2J(1+\delta^2)}-i\infty~,\\
D_+&=&1-\frac{V}{2J(1+\delta^2)}-i0^+~,
\end{eqnarray}
and the phase shifts are
\begin{equation}
\delta_-(E_i+0^+)=\pi/2~,
\end{equation}
\begin{equation}
\delta_+(E_i+0^+)=\left\{\begin{array}{lc}
0~~~V<2J(1+\delta^2)\\
\pi~~~V>2J(1+\delta^2)~.
\end{array}\right.
\end{equation}
For $E_0=E_f-0^+$, i.e., the energy is just below the top of the continuum 
band,
\begin{eqnarray}
D_-&=&1+\frac{V}{2J(1+\delta^2)}-i0^+~,\\
D_+&=&1-\frac{V}{2J(1+\delta^2)}-i\infty~,
\end{eqnarray}
and the phase shifts are
\begin{equation}
\delta_-(E_f-0^+)=0~.
\end{equation}
\begin{equation}
\delta_+(E_f-0^+)=\left\{\begin{array}{lc}
\pi/2~~~V<2J(1+\delta^2)\\
\pi/2~~~V>2J(1+\delta^2)~.
\end{array}\right.
\end{equation}

Now let us study the behavior of the phase shift as a function of energy. 
For the $A_g$ state, when the energy passes through the edge of the
band the phase shift falls discontinuously from $\pi$ to $\pi/2$, then 
gradually approaches zero at the top of the band. So there is always a bound 
state. For the $B_u$ exciton, when
$V<2J(1+\delta^2)$ the phase shift increases from zero as the energy
increases from the bottom of the band, and approaches $\pi/2$ just below the 
top
of the band, and then drops to zero again when we pass through the top edge 
of the band. Thus no bound state exists. When $V>2J(1+\delta^2)$, the phase 
shift decreases from $\pi$ to $\pi/2$ as
the energy increases from the bottom to the top of band, and
abruptly falls to zero when we cross the edge of the band. Thus a bound state 
appears. The discontinuity of
the phase shift at the band edges is due to the infinite density of
states at the bottom and top of the band.

Figure 1 shows the wave functions of the $A_g$ and $B_u$
states with $U=10t$, $V=0.5t$, and $\delta=0.2$, corresponding to 
binding energies $x_+=0.024t$ and $x_-=0.353t$, respectively. We can see
that the wave function of the $A_g$ exciton decays more rapidly than
that of the $B_u$ one. 
For $x>0$, we introduce the parameter $z$, 
\begin{equation}
z=-\ln\Bigl(\frac{1-\sqrt{1-u^2}}{u}\Bigr)>0~.
\label{width}
\end{equation}
Since for large $l$,
\begin{equation}
G(2l;E_0)\sim e^{-z|2l|}~,
\end{equation}
we can define the width $R$ of the $K=0$ exciton by $R=2/z$.
From Eq. (\ref{width}), we estimate the width
is about 3 lattice constants for the $A_g$ exciton and about 12
lattice constants for the $B_u$, as shown in Fig. 1.

We calculate the energy of the exciton for $K\ne 0$ from
$D(E_K)=0$. A straightforward computation of Eq. (\ref{gre}) gives
\begin{eqnarray}
G(0;E_K)&=&-\frac{1}{\sqrt{ac-b^2}}~,\\
G(2;E_K)&=&G^*(-2;E_K)\nonumber~,\\
&=&-\frac{1}{4b^2+(a-c)^2}\Bigl\{\bigl[2(a-c)
+\frac{c^2-a^2}{\sqrt{ac-b^2}}\bigr]
+i\bigl[4b-\frac{2b(a+c)}{\sqrt{ac-b^2}}\bigr]\Bigr\}~,
\end{eqnarray}
where
\begin{equation}
a=x+3J+5\delta^2 J+J(1-\delta^2)\cos 2K~,
\end{equation}
\begin{equation}
b=J(1-\delta^2)\sin 2K~,
\end{equation}
\begin{equation}
c=x+J(1-\delta^2)-J(1-\delta^2)\cos 2K~.
\end{equation}
By solving the equations
\begin{eqnarray}
1+G(0;E_K)V+|G(2;E_K)|V&=&0~,\\
1+G(0;E_K)V-|G(2;E_K)|V&=&0~,
\end{eqnarray}
we can compute the energy of the moving excitons. In Fig. 2, we 
have described the energy of the excitons as a function of the 
center-of-mass momentum $K$.
There are two branches in the energy spectra. The energy difference
between these two branches becomes smaller when $K$ increases, and
reaches a minimum at $K=\pi/2$. The bandwidths of these two branches are 
approximately $J$.

When $V\gg J$, our results show that both $A_g$ and $B_u$ excitons
have binding energy $V$, which is consistent with physical
intuition.\cite{GCM95} In the strong
coupling limit $U\gg V\gg t$, Guo {\it et al.} used a local (zero hopping 
limit) picture and argued, since
the ground state has all sites singly
occupied and the exciton states are linear
combinations of configuration ...11120111..., where the numbers denote
site occupancies, that the exciton energy is $U-V$. The electron-hole 
continuum consists of
all states in which the double occupancy (electron) and the empty site
(hole) are separated by more than one site (e.g., ...11211..1011...),
which has energy $U$. Thus the binding energy is $V$.
Another prediction from our theory is that the $2A_g$ state has a lower 
energy than the $1B_u$, which is the observed ordering in many 
non-luminescent conjugated polymers.\cite{BCM92,KSW88,HKS82} The strong 
Coulomb interaction regime we consider here is the reason for this ordering
in our model.

\subsection{Optical absorption}

We will calculate the optical absorption from Fermi's golden rule
\begin{equation}
\alpha(\omega)\propto \frac{1}{\omega}\sum_n|\langle n|{\bf J}|g 
\rangle |^2\delta(\omega-E_n)~,
\end{equation}
where ${\bf J}$ is the current opertor and $|n\rangle$ is the excited state 
with
one electron-hole pair. This expression can be written in a more general form
if we denote $|\nu \rangle={\bf J}|g \rangle$,
\begin{equation}
\alpha(\omega)\propto -\frac{1}{\pi\omega}\lim_{\varepsilon\to 0^+}
\Im \langle \nu|\frac{1}{\omega+i\varepsilon-H}|\nu\rangle~.
\end{equation}
Hamiltonian $H$ referred to here is that determining the energy of the 
electron-hole pair with $K=0$.
The current operator in the polymers reads
\begin{equation}
{\bf J}=it\sum_{l\sigma}[1-(-1)^l\delta](c^{\dag}_{l+1\sigma}c_{l\sigma}-
c^{\dag}_{l\sigma}c_{l+1\sigma})~.
\label{curr}
\end{equation}
Using the spinless fermions $\eta_l$ and $\gamma_l$ defined in Sec. II.A, we 
rewrite Eq. (\ref{curr}) as
\begin{equation}
{\bf J}=-it\delta\sum_{l}(\eta_{l+1}\gamma_l-\gamma^{\dag}_l\eta^{\dag}_{l+1}
+\eta_{l}\gamma_{l+1}-\gamma^{\dag}_{l+1}\eta^{\dag}_{l})~.
\end{equation}
Thus the optical absorption can be expressed by the electron-hole Green's 
function
\begin{equation}
\alpha(\omega)\propto -\frac{t^2\delta^2}{\pi\omega}\Im[\tilde{G}(0;\omega)
+\tilde{G}(2;\omega)]~,
\end{equation}
where $\tilde{G}(l;\omega)$ is the Green's function of $H$, satisfying
\begin{equation}
\tilde{\cal G}(z)={\cal G}(z)+{\cal G}(z){\cal T}(z){\cal G}(z)~.
\end{equation}
If we denote $G(n;\omega)=G_n$, then
\begin{equation}
\alpha(\omega)\propto -\frac{t^2\delta^2}{\pi\omega}\Im
\Bigl\{(G_0+G_{-2})
-\frac{V}{D(\omega)}[
(G_0+G_2)^2+V(G^3_0-G_0G^2_2+G^2_0G_2-G^3_2)]\Bigr \}
\end{equation}
and $D(\omega)=[1+(G_0-G_2)V][1+(G_0+G_2)V]$.

From Fig. 3, the $B_u$ exciton has acquired 52\% oscillator strength when 
$U=5t$ and $V=t$. If we increase $V$ and thus have an exciton with a larger 
binding energy, the $B_u$ exciton will gain more oscillator strength. For 
$U=5t$ and $V=2t$, the $B_u$ exciton
has achieved 95\% strength, and the strength of the transition to the 
continuum is correspondingly diminished, as shown in Fig. 4. 
The large transition strength for the exciton state is a
characteristic feature of one-dimension.

\section{impurities and the coherent motion of exciton}

As stated in the introduction, ``pristine'' samples of a polymer cannot 
eliminate all impurities and defects. Moreover, the fluctuations (both quantum
and thermal)
of the lattice are also a kind of intrinsic disorder for the electronic states
in polymers.\cite{MW92}
 Excitons represent a coherent composite particle motion of correlated
electrons and holes, whereas impurities tend to produce localized wavefunction.
We know that an impurity has a strong effect on transport properties in many 
systems, especially low-dimensional materials, so a natural question arises:
how do the impurities affect the coherent motion of the exciton?
 
In polymers, two kinds of impurity are often referred to in the 
literature.\cite{PBBL} 
A site impurity is represented by a local potential at site 0:
\begin{equation}
H_1=V_0\sum_{\sigma}c^{\dag}_{0\sigma}c_{0\sigma}~,
\end{equation}
and a bond impurity which acts on the bond between sites 0 and 1:
\begin{equation}
H_2=-W_0\sum_{\sigma}(c^{\dag}_{0\sigma}c_{1\sigma}+{\rm H.c.})~.
\end{equation}
Both of these impurities have very localized (on-site) potentials. However, 
for a charged impurity, its potential, in principle, may be of long-range. 
Thus we have two length scales here: one is the range of the impurity potential
($l_i$) and the other the range of the (screened) Coulomb interaction ($l_V$). 
The latter is equal to one lattice constant in our model. Since the impurity 
competes with the Coulomb interaction (in exciton states) differently in 
the two regimes ($l_V>l_i$ or $l_V<l_i$), impurity effects are expected to 
be different. 

\subsection{On-site impurity potentials}

For the site and bond impurities described above, 
we can rewrite them in the spinless fermion 
representation, giving
\begin{equation}
H_1=V_0(-\eta^{\dag}_0\eta_0+\gamma^{\dag}_0\gamma_0)~,
\end{equation}
\begin{equation}
H_2=-W_0(-\eta_0\gamma_1-\gamma^{\dag}_1\eta^{\dag}_0+\gamma^{\dag}_0
\eta^{\dag}_1
+\eta_1\gamma_0)~.
\end{equation}
Since $H_2$ involves the creation and annihilation of an electron-hole pair, 
it must be
less important than $H_1$ by order $1/U$. This can be seen more clearly by 
using the unitary transformation $H^S=e^{-S}H e^S$,
\begin{equation}
H^S_2=\frac{W^2_0}{U+2J(1-\delta^2)}(-\eta^{\dag}_0\eta_0-\eta^{\dag}_1\eta_1
-\gamma^{\dag}_0\gamma_0-\gamma^{\dag}_1\gamma_1)~.
\end{equation}
Although the site impurity seems more realistic from the above analysis, we 
will study 
three kinds of impurity to arrive at a unified picture of impurity effects:
\begin{eqnarray}
H^{\rm imp}_1&=&V_0(-\eta^{\dag}_0\eta_0+\gamma^{\dag}_0\gamma_0),\\
H^{\rm imp}_2&=&V_0(-\eta^{\dag}_0\eta_0-\gamma^{\dag}_0\gamma_0),\\
H^{\rm imp}_3&=&V_0(\eta^{\dag}_0\eta_0+\gamma^{\dag}_0\gamma_0).
\end{eqnarray}
Hamiltonian $H^{\rm imp}_1$, in which the impurity attracts the hole and 
repels the 
electron, or vice versa, imitates a local charged impurity. Hamiltonian 
$H^{\rm imp}_2$, in which the impurity
attracts both the electron and hole, acts as a trap for particles. In 
Hamiltonian $H^{\rm imp}_3$, the 
impurity potentials are repulsive for both the electron and hole, describing a 
barrier effect. The last two types of impurity can be viewed as simulating 
the cross-linking and
conjugation breaking effects in conjugated polymers.

There is no translation invariance once the impurity is included, so we will 
work in real space and the Hamiltonian we must study reads
\begin{eqnarray}
H_i&=&\sum_l\bigl\{J(1+\delta^2)\eta^{\dag}_l\eta_l+\frac{1}{2}J(1-\delta^2)
(\eta^{\dag}_l\eta_{l+2}
+\eta^{\dag}_{l+2}\eta_l)+[U+J(1-3\delta^2)]\gamma^{\dag}_l\gamma_l\nonumber\\
&+&\frac{1}{2}J(1-\delta^2)(\gamma^{\dag}_l\gamma_{l+2}
+\gamma^{\dag}_{l+2}\gamma_l)\bigr\}
+V\sum_l(\eta^{\dag}_{l+1}\eta^{\dag}_l\eta_l\eta_{l+1}+
\gamma^{\dag}_{l+1}\gamma^{\dag}_l\gamma_l\gamma_{l+1}\nonumber\\
&-&\eta^{\dag}_{l+1}\gamma^{\dag}_l\gamma_l\eta_{l+1}
-\gamma^{\dag}_{l+1}\eta^{\dag}_l\eta_l\gamma_{l+1})
+H^{\rm imp}_i~.
\end{eqnarray}

The key issue here is how to measure the coherence in the excitonic
composite particle. We can do this by 
defining the correlation function between the electron and hole in the lowest 
state in the one electron and one hole subspace:
\begin{equation}
{\cal R}(l,l')=\frac{\langle\delta\rho_h(l)\delta\rho_e(l')\rangle}
{\sqrt{\langle(\delta\rho_h(l))^2\rangle \langle(\delta\rho_e(l'))^2\rangle}}~,
\label{cor}
\end{equation}
where the deviations are
\begin{equation}
\delta A=A-\langle A \rangle~,
\end{equation}
and the density operators of electron and hole are
\begin{eqnarray}
\rho_h(l)&=&\eta^{\dag}_l\eta_l~,\\
\rho_e(l)&=&\gamma^{\dag}_l\gamma_l~.
\end{eqnarray}

In the impurity-free system, this correlation function (\ref{cor}) 
is connected with the 
relative wavefunction of the lowest exciton state 
$|\Psi_0 \rangle=\frac{1}{\sqrt{N}}\sum_sB_s\gamma^{\dag}_{l+s}\eta^{\dag}_l|
g\rangle$
by
\begin{equation}
{\cal R}^{\rm free}(l,l')=|B_{l-l'}|^2~.
\end{equation}

When we add an impurity, the correlation is expected to decrease. The 
closer ${\cal R}$ is to ${\cal R}^{\rm free}$, the more correlated are the 
electron and hole in the lowest excitonic
states, while ${\cal R}$ approaching zero means that there is no correlation 
between the electron and 
hole; in other words, this excitonic state has lost all its coherence. 

The effects of the first kind of impurity are illustrated in Fig. 5, which 
shows 
the electron-hole correlation functions for different sites in a finite system
of size $N=10$.
We emphasize that the parameters we use ($U=10t$ and $V=t$) ensure that the 
exciton has a 
very localized {\em relative} wavefunction, so that finite system corrections 
and the boundary condition effects are not important. In Fig. 5, all the 
correlation functions
exhibit a crossover behavior around $V_0\sim 0.1t$. This can be understood in 
term of a relevant energy
scale of the exciton, namely, the width of the exciton band, which equals 
$J=2t^2/U$. This 
crossover behavior, which occurs at $V_0\sim J$, describes the localization of
the exciton, i.e., the free exciton 
becomes trapped. We can also calculate the charge density at the impurity 
site as the 
impurity strength increases. Since in an impurity-free system the hole 
(electron) is uniformly
distributed, and from our exciton theory the electron and hole do not tend to
occupy the same site, the hole density at the impurity site is $2/N$. We see a 
crossover again in Fig. 6 when $V_0$ is comparable to the bandwidth $J$. 
After this crossover, the hole density at the impurity site approaches 1,
clearly showing that
the exciton is trapped by the impurity, and the correlation between 
the electron and hole gradually vanishes, although, as indicated in Fig. 7, 
they are bound together near the impurity.

For the second type of impurity, from Fig. 8,
a crossover is also observed if the impurity strength is similar
to the exciton bandwidth, again indicating that the exciton is trapped. But 
when $V_0$ is larger
than $V$, the correlation function abruptly falls to zero, which implies 
the total breakdown of the exciton as a composite particle. This is because, 
when $V_0$ is large enough, 
it is a lower energy for the impurity to trap the electron and hole separately 
rather than the impurity trapping
the hole and then the electron being trapped near the hole due to the Coulomb 
interaction (as for
the first kind of impurity). Thus the electron and the hole occupy the same 
site and they have no Coulomb interaction. This is not an exciton.

Now we consider the third type of impurity. The correlation function 
behaviors in Fig. 9 seem more complicated   
than for the other two types of impurity. The correlation function in which 
the hole is at the impurity site
shows an analogous crossover behavior when $V_0$ is near the exciton bandwidth
$J$ to the first
and second types of impurity. However, if both the electron and the hole are 
left (or right) of the
impurity, from the correlation function we see that they have not felt the 
impurity. On the other hand, if
the electron and the hole sit on different sides of the impurity, there is a 
crossover at $V_0\sim J$,
but part of correlation between the electron and hole survives. 

The different effects of these three kinds of impurity can be further 
understood if 
we project the lowest excitonic state to the free exciton states with momentum
$K$. We depict in Figs. 10, 11 and 12 the 
distribution $|Z_K|^2$, where $Z_K=\langle \Psi_K|\Psi \rangle$, and 
$|\Psi\rangle$
is the lowest excitonic state in these three disordered systems.
In the impurity-free system, the lowest
state is the linear combination of exciton states with $K=0$ and $K=\pi$ 
(they are degenerate). In
the presence of an impurity, the exciton state will be scattered to other 
exciton states with
different $K$ and the distribution of momenta will broaden from the $\delta$ 
function. A free 
exciton state with a specific $K$ can be defined only when the width of the 
distribution of momentum
is narrow enough (i.e., the lifetime of this state is long enough). This is 
analogous to 
the quasi-particle in Landau's Fermi Liquid theory.\cite{AGD}
 From these figures, we see that for the first and 
second types of impurities, after the crossover at $V_0\sim J$ the distribution
in momentum space 
becomes so broad that we can hardly identify the original exciton state with
momentum $K=0$ or $K=\pi$.
For the second type of impurity, when $V_0$ is larger than $V$, the final 
state has no distribution 
at all on any free exciton state, also indicating that the final state is no 
longer excitonic.
However, for the third kind of impurity, after the crossover at $V_0\sim J$, 
the distribution
in momentum space still has two sharp peaks at $K=0$ and $K=\pi$. This is the
reason why the 
exciton is still coherent, as shown in the correlation functions, and in this 
sense the exciton can be 
regarded as a quasi-particle in this disordered system.

\subsection{Extended impurity potentials}

For a charged impurity, the range of (screened Coulomb) potential can be extended 
over several 
lattice constants. As an illustration, here we consider a specific impurity potential
\begin{equation}
H^{\rm imp}_4=V_0(-\eta^{\dag}_0\eta_0+\gamma^{\dag}_0\gamma_0)+\frac{V_0}{2}
(-\eta^{\dag}_1\eta_1+\gamma^{\dag}_1\gamma_1
-\eta^{\dag}_{-1}\eta_{-1}+\gamma^{\dag}_{-1}\gamma_{-1})~.
\end{equation}
Its range ($l_i$) is three lattice constants which is longer than that of the 
Coulomb interaction ($l_V$). The correlation functions are illustrated in 
Fig. 13, 
from which we observe again a crossover around $V_0\sim J$, indicating the 
free exciton becomes trapped. Interestingly, when $V_0$ is sufficiently large 
compared to $V$, the correlation functions falls further abruptly to zero, 
indicating 
the dissociation of the exciton into an uncorrelated electron-hole pair. Note 
that this does not occur for the charged impurity $H^{\rm imp}_1$ with the 
on-site potential, since the impurity range is then less than the trapped 
exciton size.
From charge densities shown in Fig. 14, we find that for 
$V_0=0.5t$ (just after the first crossover), both the electron and the hole 
(thus the exciton) are trapped around the impurity.  For $V_0=5t$ (after the 
correlation goes to zero), the hole is trapped by the impurity while the 
electron is repelled from the impurity. 
The dissociation of excitons here is easily understood.  Because the impurity 
attracts the hole and repels the electron, when the impurity strength becomes 
sufficiently strong, the Coulomb attraction cannot overcome the impurity 
repulsion to bind the electron and hole together. This impurity-induced exciton 
dissociation may be invoked to interpret impurity-enhanced photoconductivity 
observed in certain experiments.  We can project the lowest excitonic state in 
the system with the impurity to free exciton states with different momenta. 
In Fig. 15, the distribution in momentum space changes from a sharply localized
one ($V_0=0.1t$, before the crossover) to a very broad Gaussian 
distribution ($V_0=0.5t$, after the crossover) and finally goes to zero
($V_0=5t$).  This is consistent with the picture that the free exciton becomes 
trapped, then dissociates into an uncorrelated electron-hole pair with 
increasing impurity strength.

\section{interchain coupling and interchain excitons}

The interchain coupling can strongly influence the energy and 
stability of the nonlinear excitations such as solitons and polarons.
Current interest in interchain coupling and the intrachain and interchain 
exciton 
crossover in polymers stems from the experimentally observed large amount of
interchain excitations in luminescent polymers like PPV.
However,
the concept of an interchain exciton and how to distinguish the interchain and
intrachain 
excitons, are not very clear. The wavefunction is not so useful to
specify whether a state is an interchain or intrachain exciton, because the 
wavefunction of any state, 
in principle, will spread over the whole system if interchain coupling is 
present.

To demonstrate interchain exciton states in our approach, we study a two-chain
system coupled by the nearest-neighbor hopping, 
\begin{equation}
H_{\rm hop}=-t_{\perp}\sum_{l\sigma}(c^{\dag}_{1l\sigma}c_{2l\sigma}+{\rm H.c.}
)~,
\end{equation}
and nearest-neighbor interchain Coulomb interaction,
\begin{equation}
H_{\rm Cou}=V_{\perp}\sum_l(\rho_{1l}-1)(\rho_{2l}-1)~.
\end{equation}
Here $1,2$ are chain indices.
Now the unperturbed Hamiltonian in the large-$U$ limit becomes
\begin{eqnarray}
\tilde{H}_0&=&H^{\prime}_0+H_{\rm hop}\nonumber\\
&=&\sum_{i,k}(\epsilon_k\eta^{\dag}_{ik}\eta_{ik}
+\tilde{\epsilon}_k\gamma^{\dag}_{ik}\gamma_{ik})+t_{\perp}\sum_l
(\eta^{\dag}_{2l}\eta_{1l}-\gamma^{\dag}_{1l}\gamma_{2l}+{\rm H.c.})~.
\end{eqnarray}
In momentum space, this is
\begin{equation}
\tilde{H}_0=\sum_{i,k}(\epsilon_k\eta^{\dag}_{ik}\eta_{ik}
+\tilde{\epsilon}_k\gamma^{\dag}_{ik}\gamma_{ik})+t_{\perp}\sum_k
(\eta^{\dag}_{2k}\eta_{1k}-\gamma^{\dag}_{1k}\gamma_{2k}+{\rm H.c.})~,
\label{inter}
\label{INTER}
\end{equation}
which can be readily diagonalized, yielding
\begin{equation}
\tilde{H}_0=\sum_{I,k}(E^I_{k}\tilde{\eta}^{\dag}_{Ik}\tilde{\eta}_{Ik}
+\tilde{E}^I_k\tilde{\gamma}^{\dag}_{Ik}\tilde{\gamma}_{Ik})~.
\end{equation}
Here $I=1,2$ is the band index (hereafter we use small $i,j$ for the chain 
indices, and capital
$I,J$ for the band indices). The details of the diagonalization are given in 
Appendix B.
Now we have two conduction and two valence bands with the dispersion relations
\begin{eqnarray}
E^{1,2}_k&=&\epsilon_k\mp t_{\perp}~,\\
\tilde{E}^{1,2}_k&=&\tilde{\epsilon}_k\mp t_{\perp}~.
\end{eqnarray}
The interaction Hamiltonian now contains two parts
\begin{equation}
H_V=H_{\rm int}+H_{\rm Cou}~.
\end{equation}
The electron-hole interaction part of $H_V$, which is relevant to the exciton 
state, is
\begin{equation}
H^{\rm e-h}_V=-V\sum_{i,l}(\eta^{\dag}_{il+1}\gamma^{\dag}_{il}\gamma_{il}
\eta_{il+1}
+\gamma^{\dag}_{il+1}\eta^{\dag}_{il}\eta_{il}\gamma_{il+1})
-V_{\perp}\sum_l(\eta^{\dag}_{1l}\gamma^{\dag}_{2l}\gamma_{2l}\eta_{1l}
+\gamma^{\dag}_{1l}\eta^{\dag}_{2l}\eta_{2l}\gamma_{1l})~.
\label{hveh}
\end{equation}
If we define local fermion operators
\begin{eqnarray}
\tilde{\eta}_{Il}&=&\frac{1}{\sqrt{N}}\sum_k e^{ikl}\tilde{\eta}_{Ik}~,\\
\tilde{\gamma}_{Il}&=&\frac{1}{\sqrt{N}}\sum_k e^{ikl}\tilde{\gamma}_{Ik}~,
\end{eqnarray}
we can rewrite Eq. (\ref{hveh}) as
\begin{eqnarray}
H^{\rm e-h}_V&=&-\frac{V}{2}\sum_{IJ}\sum_l(\tilde{\eta}^{\dag}_{Il+1}
\tilde{\gamma}^{\dag}
_{Jl}\tilde{\gamma}_{Jl}\tilde{\eta}_{Il+1}
+\tilde{\gamma}^{\dag}_{Il+1}\tilde{\eta}^{\dag}
_{Jl}\tilde{\eta}_{Jl}\tilde{\gamma}_{Il+1})\nonumber\\
&+&\frac{V}{2}\sum_l(\tilde{\eta}^{\dag}_{1l+1}\tilde{\gamma}^{\dag}
_{1l}\tilde{\gamma}_{2l}\tilde{\eta}_{2l+1}
+\tilde{\eta}^{\dag}_{1l+1}\tilde{\gamma}^{\dag}
_{2l}\tilde{\gamma}_{1l}\tilde{\eta}_{2l+1}+
\tilde{\gamma}^{\dag}_{1l+1}\tilde{\eta}^{\dag}
_{1l}\tilde{\eta}_{2l}\tilde{\gamma}_{2l+1}
+\tilde{\gamma}^{\dag}_{1l+1}\tilde{\eta}^{\dag}
_{2l}\tilde{\eta}_{1l}\tilde{\gamma}_{2l+1}+1\Longleftrightarrow 2)\nonumber\\
&-&\frac{V_{\perp}}{2}\sum_{IJ}\sum_l\tilde{\gamma}^{\dag}_{Il}
\tilde{\eta}^{\dag}
_{Jl}\tilde{\eta}_{Jl}\tilde{\gamma}_{Il}
-\frac{V_{\perp}}{2}\sum_l(\tilde{\eta}^{\dag}_{1l}\tilde{\gamma}^{\dag}
_{1l}\tilde{\gamma}_{2l}\tilde{\eta}_{2l}
+\tilde{\eta}^{\dag}_{1l}\tilde{\gamma}^{\dag}
_{2l}\tilde{\gamma}_{1l}\tilde{\eta}_{2l}+1\Longleftrightarrow 2)~.
\end{eqnarray}
Thus we can construct the exciton wavefunction in this two-chain system as
\begin{equation}
|\Psi_K\rangle=\sum_{IJ}\sum_s B^{IJ}_{s,K}|\psi^{IJ}_{s,K}\rangle~.
\end{equation}
Here the center-of-mass momentum $K$ is still a good quantum number, and the 
basis is
\begin{equation}
|\psi^{IJ}_{s,K}\rangle=\frac{1}{\sqrt{N}}\sum_le^{iKl}
\tilde{\gamma}^{\dag}_{Il+s}
\tilde{\eta}^{\dag}_{Jl}|g\rangle~,
\label{bas2}
\end{equation}
which means the exciton state is a combination of every possible electron-hole
excitation in different conduction and valence bands.
The free electron-hole pair Green's function under the basis Eq. (\ref{bas2}) is
\begin{equation}
\langle\psi^{I'J'}_{r,K}|\frac{1}{z_K-\tilde{H}_0}|\psi^{IJ}_{s,K}\rangle
=\delta_{II'}\delta_{JJ'}G^{IJ}(r-s;z_K)
\end{equation}
\begin{equation}
G^{IJ}(l;z_K)=\frac{1}{N}\sum_k\frac{e^{ikl}}{z_K-(\tilde{E}^I_k+E^J_{-k+K})}~,
\end{equation}
and the scattering potential is written as
\begin{equation}
\langle\psi^{I'J'}_{s,K}|H^{\rm e-h}_V|\psi^{IJ}_{s',K}\rangle
=\delta_{ss'}\Bigl(-\frac{V}{2}\delta_{s1}-\frac{V}{2}\delta_{s-1}-
\frac{V_{\perp}}{2}
\delta_{s0}\Bigr)F(I',J';I,J)
\end{equation}
with $F(I,J;I,J)=1$ and $F(1,1;2,2)=F(2,2;1,1)=F(1,2;2,1)=F(2,1;1,2)=1$, and 
otherwise
$F=0$. We can solve for the exciton states by locating the roots of the 
determinant
$\det (1-{\cal GU})$ according to $t$ matrix theory. The whole determinant 
can be
decomposed into blocks by appropriate transformations and we achieve two 
subdeterminants,
\begin{equation}
D_1=\left| \begin{array}{cc}
1+\displaystyle\frac{V_{\perp}}{2}G^{11}(0;z_K)  
& \displaystyle\frac{V_{\perp}}{2}G^{11}(0;z_K)\\
\displaystyle\frac{V_{\perp}}{2}G^{22}(0;z_K) & 
1+\displaystyle\frac{V_{\perp}}{2}G^{22}(0;z_K)
\end{array} \right|
\end{equation}
and
\begin{equation}
D_2=\left| \begin{array}{cc}
1+\displaystyle\frac{V_{\perp}}{2}G^{12}(0;z_K) 
 & \displaystyle\frac{V_{\perp}}{2}G^{12}(0;z_K)\\
\displaystyle\frac{V_{\perp}}{2}G^{21}(0;z_K) 
& 1+\displaystyle\frac{V_{\perp}}{2}G^{21}(0;z_K)
\end{array} \right|~.
\end{equation}
$D_1$ and $D_2$ are determined only by interchain Coulomb interaction 
$V_{\perp}$ corresponding to {\em interchain} excitons,
and two subdeterminants,
\begin{equation}
D_3=\left| \begin{array}{cc}
1+\displaystyle\frac{V}{2}[G^{11}(0;z_K)+G^{22}(0;z_K)]
& \displaystyle\frac{V}{2}[G^{11}(-2;z_K)+G^{22}(-2;z_K)]\\
\displaystyle\frac{V}{2}[G^{11}(2;z_K)+G^{22}(2;z_K)] & 
1+\displaystyle\frac{V}{2}[G^{11}(0;z_K)+G^{22}(0;z_K)]
\end{array} \right|
\end{equation}
and
\begin{equation}
D_4=\left| \begin{array}{cc}
1+\displaystyle\frac{V}{2}[G^{12}(0;z_K)+G^{21}(0;z_K)]
& \displaystyle\frac{V}{2}[G^{12}(-2;z_K)+G^{21}(-2;z_K)]\\
\displaystyle\frac{V}{2}[G^{12}(2;z_K)+G^{21}(2;z_K)] & 
1+\displaystyle\frac{V}{2}[G^{12}(0;z_K)+G^{21}(0;z_K)]
\end{array} \right|~,
\end{equation}
which are determined only by the intrachain Coulomb interaction $V$ 
corresponding 
to {\em intrachain} excitons. Equations $D_1=0$ and $D_2=0$ have a single root,
respectively,
whereas both $D_3=0$ and $D_4=0$ have two roots. Thus there are a total of 
six exciton 
bands: two interchain exciton bands and four intrachain exciton bands.
Figures 16 and 17 show these intrachain and interchain exciton bands. The 
relative energy ordering of the interchain and intrachain excitons 
depends on the ratio $V_{\perp}/V$. 

It is interesting to study the wave function of the interchain excitons. The 
static interchain exciton can be represented in real space as
\begin{equation}
|\Psi_{K=0}\rangle=\sum_{ij}\sum_s A^{ij}_s\frac{1}{\sqrt{N}}\sum_l
\gamma^{\dag}_{il+s}\eta^{\dag}_{jl}|g\rangle~.
\end{equation}
For the lower exciton state determined by $D_1(z_0)=0$, we obtain
\begin{equation}
A^{11}_s=-A^{22}_s=-\frac{G^{11}(s;z_0)-G^{22}(s;z_0)}{2\sqrt{-[G^{11}(0;z_0)
+G^{22}(0;z_0)]'}}~,
\end{equation}
which represents the amplitude for the electron and hole being on the same 
chain; while
\begin{equation}
A^{21}_s=-A^{12}_s=-\frac{G^{11}(s;z_0)+G^{22}(s;z_0)}{2\sqrt{-[G^{11}(0;z_0)
+G^{22}(0;z_0)]'}}
\end{equation}
is the amplitude that the electron and hole are on different chains. Here
\begin{equation}
G^{11}(0;z)=-\frac{1}{\sqrt{(E_G-z-2t_{\perp})(E_G-z-2t_{\perp}+W)}}
\end{equation}
\begin{equation}
G^{22}(0;z)=-\frac{1}{\sqrt{(E_G-z+2t_{\perp})(E_G-z+2t_{\perp}+W)}}~,
\end{equation}
where $E_G$ is defined as Eq. (\ref{gap}) and $W=4J(1+\delta^2)$.
Figure 18 illustrates the intrachain wavefunction $A^{11}_s$ and interchain 
wavefunction $A^{21}_s$ for this interchain exciton state. We can see that 
although the state is for an interchain exciton,
there is still some probability for the electron and hole to be on the same 
chain.

For the higher interchain exciton determined by $D_2(z_0)=0$, we have
\begin{equation}
A^{11}_s=A^{22}_s=0~,
\end{equation}
\begin{equation}
A^{21}_s=A^{12}_s=-\frac{G^{12}(s;z_0)}{\sqrt{-[G^{12}(0;z_0)]'}}~,
\end{equation}
\begin{equation}
G^{12}(0;z)=-\frac{1}{\sqrt{(E_G-z)(E_G-z+W)}}~.
\end{equation}
In this interchain exciton, there is no relative amplitude between the 
electron and hole 
if they are on the same chain. In Fig. 19, we plot the wavefunction $A^{21}_s$ 
for this interchain exciton.

For more complicated Coulomb interactions within and between the chains, the 
exact 
interchain and intrachain exciton poles are difficult to obtain analytically.
Instead, we can measure the correlation between the electron and hole using 
Eq. (\ref{cor}). In Fig. 20, we depict the intrachain and interchain 
electron-hole correlation functions for our simple interchain coupling 
situation for two $N=12$ chains. Here we choose a fixed interchain hopping
$t_{\perp}/t=0.2$.
The transition at $V_{\perp}/V=1.18$ shows that the
lowest exciton state changes from an intrachain exciton to an interchain one.

\section{charge transfer in a molecularly-doped polymer}

Photoinduced charge transfer from a polymer chain to an adjacent dopant 
molecule, such as in PPV/C$_{60}$ blends has attracted much recent attention, 
because it
can greatly increase the photoconductivity in polymers. In recent theoretical
work on this phenomenon, Rice and Gartstein\cite{RG96} proposed a mechanism to 
explain the observed ultrafast time scale
of this process. In this section, instead of discussing the time scale, we 
attempt
to calculate the final state wavefunction of the whole system comprising the 
polymer chain and dopant molecule. This can tell us what part of the electron 
in the exciton has transferred from the chain to the 
dopant. To make our idea more transparent, let us briefly describe this 
photoinduced
charge transfer process. In the ground state, there is no overlap between the 
chain and 
the dopant. The photoexcitation produces an exciton state in the polymer chain.
Then, due to the coupling between
the polymer and the dopant molecule, the electron (or hole) will transfer from 
the chain to the adjacent molecule. 
As a simplified Hamiltonian, we consider 
\begin{equation}
H=H_{\rm chain}+\Delta_e\sum_mc^{\dag}_mc_m+H_{\rm tran}~.
\end{equation}
Here we are modeling the dopant molecule by assuming it has an acceptor level 
with energy 
$\Delta_e$, which couples to the polymer chain only by nearest-neighbor 
hopping 
\begin{equation}
H_{\rm tran}=-v\sum_m(c^{\dag}_m\gamma_0+{\rm H.c.})
=-\frac{v}{\sqrt{N}}\sum_{m,k}(c^{\dag}_m\gamma_k+{\rm H.c.})~.
\end{equation}
A schematic diagram of our model is shown in Fig. 21.
The whole system consists of a polymer chain and $N_d$ dilute noninteracting 
dopants. Before the 
coupling between the polymer chain and the dopant is switched on, the system 
has an
exciton state on the polymer chain. When we turn on the coupling, the electron 
in the exciton will transfer
between the chain and the molecule. Thus we can construct a variational 
wavefunction
\begin{equation}
|\Psi\rangle=a|\Psi_0\rangle +\frac{1}{\sqrt{N_d}}\sum_m\sum_k
a_kc^{\dag}_m\gamma_k|\Psi_0\rangle~,
\end{equation}
with the condition $\langle \Psi|\Psi \rangle=1$. The first term describes the 
electron remaining on the chain as a component of the exciton and the 
second term
describes that the electron with different momentum has different probability 
to transfer
to the dopant molecule. Here $|\Psi_0 \rangle$ is the assumed static exciton 
state within the polymer chain,
\begin{equation}
|\Psi_0 \rangle=\sum_s B_s\frac{1}{\sqrt{N}}\sum_l
\gamma^{\dag}_{l+s}\eta^{\dag}_l|g\rangle~,
\end{equation}
which can be represented by the relative momentum between the electron and 
hole 
\begin{equation}
|\Psi_0 \rangle=\sum_kB_k\frac{1}{\sqrt{N}}\gamma^{\dag}_k\eta^{\dag}_{-k}|
g\rangle~.
\end{equation}
Its energy is
\begin{equation}
\langle \Psi_0|H_{\rm chain}|\Psi_0\rangle=E_0~.
\end{equation}
The variational state $|\Psi\rangle$ must have
a lower energy than $E_0$~,
\begin{equation}
\epsilon=\langle\Psi|H|\Psi \rangle-E_0<0~.
\end{equation}
From $\displaystyle\frac{\partial\epsilon}{\partial a}=0$ and 
$\displaystyle\frac{\partial\epsilon}{\partial a_k}=0$, we obtain two 
coupled equations:
\begin{eqnarray}
-\sqrt{\frac{N_d}{N}}a v&=&a_k[\epsilon+E_0-(\epsilon_k+\Delta_e)]~,
\label{eq1}\\
a\epsilon&=&\frac{1}{N}\sum_k a_k|B_k|^2\sqrt{\frac{N_d}{N}} v~.                   
\label{eq2}
\end{eqnarray}
Using Eq. (\ref{eq1}), $a_k$ can be eliminated from Eq. (\ref{eq2}) 
and we have the eigenvalue equation for $\epsilon$
\begin{equation}
\epsilon=c\frac{1}{N}\sum_k \frac{v^2|B_k|^2}{\epsilon+E_0-(\epsilon_k+
\Delta_e)}\equiv c F(\epsilon)~.
\end{equation}
Here $c\equiv N_d/N$ is the dopant concentration. Once we have found the 
negative solution of $\epsilon$, 
the probability that the exciton remains on the chain is
\begin{equation}
a^2=\frac{1}{1-c F'(\epsilon)}~,
\end{equation}
where $F'(\epsilon)=\frac{dF(\epsilon)}
{d\epsilon}$. So the probability of charge transfer is
\begin{equation}
P=1-a^2=\frac{-cF'(\epsilon)}{1-cF'(\epsilon)}~.
\end{equation}

For the $B_u$ exciton in the polymer chain
\begin{equation}
B_k=\sum_lB^+_le^{-ikl}~.
\end{equation}
If we assume for demonstration purposes that the exciton is highly localized 
(this is not necessary in our theory),
i.e., $B^+_1=B^+_{-1}=\frac{1}{\sqrt{2}}$ and $B^+_l=0$ for other $l$, then
\begin{equation}
B_k=\sqrt{2}\cos k~,
\end{equation}
and we can write $F(\epsilon)$ in a very compact form:
\begin{equation}
F(\epsilon)=\frac{v^2}{J(1-\delta^2)}\Bigl(\sqrt{\frac{E+2J(1-\delta^2)}{E}}
-1\Bigr)~,
\end{equation}
with
\begin{equation}
E=-(\epsilon+E_0-\Delta_e-2J\delta^2)~.
\end{equation}

Figure 22 illustrates the probability of charge transfer to the dopant. The 
dopant
concentration is set to be 0.2, a typical value for a molecularly-doped 
polymer. We see that
when the acceptor level is near the exciton energy $E_0$, a crossover will
occur. When the acceptor level is below this crossover value, the electron is 
mainly on the 
dopant. Otherwise, the electron is mainly on the polymer chain. The coupling 
strength
$v$ affects the charge transfer by controlling the width of the crossover. 
The smaller $v$ is, the more rapid is the crossover.

For the $A_g$ exciton, if we make the assumption $B^-_1=-B^-_{-1}=\frac{1}
{\sqrt{2}}$, then
\begin{equation}
|B_k|=|\sum_lB^-_le^{ikl}|=\sqrt{2}\sin k~,
\end{equation}
and we have
\begin{equation}
F(\epsilon)=\frac{v^2}{J(1-\delta^2)}\Bigl(1-\sqrt{\frac{E}{E+2J(1-\delta^2)}}
\Bigr)~.
\end{equation}
We plot the charge transfer probability for the $A_g$ state in Fig. 23. We 
can see that
there is a threshold for $\Delta_e$. Below this value, the electron will
thoroughly transfer
to the dopant molecule. However for the $B_u$ state there is a long tail below
the critical value, indicating
some fraction of the electron can still be found in the polymer. Having gained 
the knowledge of how $\Delta_e$, $v$, and the exciton wavefunction influence 
the
probability of charge transfer, one will be able to control this transfer 
process in the conjugated polymer.

\section{concluding remarks}

In this paper, we have extensively studied the exciton states in conjugated 
polymers
by emphasizing the dominant role of {\em e-e} correlations. The model we 
studied here is the
widely-used Peierls-extended Hubbard model with frozen bond dimerization. 
First, in the large-$U$ approximation,
we mapped this model to a spinless fermion model with only
nearest-neighbor Coulomb interaction in real space. The short range 
interaction 
enabled us to apply $t$ matrix theory to analytically calculate the energy 
spectrum and
wavefunction of bound (exciton) states. We have found that there always exists
a stable $A_g$ exciton as long as the nearest-neighbor Coulomb $V$ is nonzero;
for the $B_u$ state, however, a stable exciton state can exist only when $V$ 
is larger 
than the half width of the continuum band. This criterion has been proven
based on Levinson's theorem. In our results, we have a 
correct ordering for $A_g$ and $B_u$ states, i.e., $2A_g<1B_u$, as observed 
in most conjugated
polymers.
The impurity effects on the coherent motion of the 
excitons were also investigated in this large-$U$ approximation. The coherence
of the exciton 
can be measured by an appropriately defined electron-hole correlation function.
We have
studied impurities with on-site potentials as well as a charged impurity with
a more extended potential. There are 
three kinds of impurity with the on-site potential: 
the first is like a local charge, attracting
holes but repelling electrons (or vice versa); the second acts as a well, 
attracting
both electron and hole; the third is like a barrier which repels both 
electron and 
hole. We have found that for the first and second type of impurities, the 
electron-hole correlations exhibit a crossover when the impurity strength
$V_0$ is
comparable to the exciton bandwidth $J$, which describes the exciton being 
trapped by 
the impurity. For the second type of impurity, if the impurity strength is 
larger than
the Coulomb interaction $V$, the deep well will trap the electron and hole 
separately, leading to the total de-correlation
of the exciton as a particle. For the third type of impurity,
the exciton coherence can survive the impurity and the distribution in
momentum space
has a sharp peak which means the exciton  still moves freely. For the charged
impurity with an extended potential of range greater than $l_V$, the range of
the Coulomb interaction, the free exciton becomes trapped at
$V_0\sim J$, analogous to the situation for on-site impurity potentials. 
However, unlike the charged impurity with the on-site potential, 
the exciton dissociates into an uncorrelated electron-hole pair
when $V_0$ is sufficiently large compared to the Coulomb strength $V$. 
 
We have also investigated the effects of interchain coupling and the resulting
interchain exciton
states within the strong correlation approximation by considering a two-chain 
system with nearest-neighbor interchain hopping $t_{\perp}$ and Coulomb
interaction $V_{\perp}$. In this coupled system, we have two conduction bands 
and 
two valence bands. Within the $t$ matrix formalism, we have found six poles for
every
center-of-mass momentum $K$, in which two poles are determined solely by 
$V_{\perp}$,
corresponding to interchain excitons, while the other four poles are  
determined solely by the intrachain Coulomb interaction $V$, corresponding to 
intrachain excitons. We have also illustrated the wavefunctions of the static 
interchain exciton. There is still some amplitude for the electron and hole 
being on 
the same chain for the interchain exciton state. For more complicated
Coulomb potentials, we propose a way to distinguish the interchain and 
intrachain 
excitons, namely by comparing the interchain electron-hole correlation 
function with the intrachain one.

The charge transfer in a molecularly-doped conjugated polymer has been studied by
constructing a variational wavefunction for the whole system including the 
polymer
chain and dopant molecule. We modeled this coupled system by simply regarding
the 
molecule has an acceptor level which interacts with the polymer chain by 
nearest-neighbor hopping $v$.
Minimizing the energy of the state, we have obtained the
energy of the variational state and, accordingly, the probability of charge
transfer. 
We have shown that a crossover behavior will occur when the acceptor level is 
near 
to the exciton energy. When the acceptor level is higher than this crossover
value,
the electron mainly remains on the polymer chain. Otherwise, most of the
electron density will transfer to the dopant molecule. 
The hopping $v$ controls the width of this crossover, the larger $v$ is, the 
more gentle is the crossover. The wavefunction is also an important influence 
on the charge
transfer. For the $A_g$ state, there is a threshold for the acceptor level. 
If $\Delta_e$
is less than this value, the charge transfers to the molecule thoroughly and 
the percentage of electron density in the polymer chain is zero.

Our calculations in this paper presented a comprehensive picture of the 
exciton in 
conjugated polymers, in a limit in which the electron correlation effects have
been taken seriously and consistently.
Our exciton theory can be readily extended to a system with a relatively
long range Coulomb interaction. Also using our spinless fermion representation
for
the Peierls-extended Hubbard model, biexciton states can be obtained either by
the Heitler-London method or diagonalization of the Hamiltonian in two
electron-hole 
pair space. Although for real conjugated polymers, the Hubbard $U$ is not so 
strong
and our results cannot quantitatively match the energy levels in luminescent 
polymers, 
our theory is useful for understanding several puzzles which have arisen from 
correlation effects 
in conjugated polymers. Finally, we note that recent experimental evidence has 
demonstrated that 
there is an excitonic contribution to the pairing mechanism in 
YBa$_2$Cu$_3$O$_{7-\delta}$.\cite{Ho94} 
We expect that our exciton theory can give some 
guidance for exciton effects in high-T$_c$ superconductors by
extending the formalism to two dimensions. 

\acknowledgments

We are grateful to X. Sun, Z. Shuai, W. Z. Wang, J. T. Gammel, S. A. Brazovskii, 
N. N. Kirova,
and D. Schmeltzer for many helpful discussions and criticisms.
This work was supported by the U. S. Department of Energy.

\appendix
\section{Explicit expressions of Green's functions}

In this Appendix, we will give explicit expressions for $G(0;E_0)$ and 
$G(2;E_0)$ in different energy regions, calculated according to the definition 
Eq. (\ref{gre}).
Here we use the same notation as in the text; $x=E_G-E_0$.
When $E_0>E_f$, i.e., $x<-4J(1+\delta^2)$, the energy is above the top of the 
continuum band, and the Green's functions are
\begin{equation}
G(0;E_0)=\frac{1}{\sqrt{x^2-4J(1+\delta^2)|x|}}~,
\end{equation}
\begin{equation}
G(2;E_0)=-\frac{1}{2J(1+\delta^2)}+\Bigl[\frac{|x|}{2J(1+\delta^2)}-
1 \Bigr]
\frac{1}{\sqrt{x^2-4J(1+\delta^2)|x|}}~.
\end{equation}
When $E_i<E_0<E_f$, i.e., $-4J(1+\delta^2)<x<0$, the energy is within the 
continuum band,
\begin{equation}
G(0;E_0)=-i\frac{1}{\sqrt{-x^2-4J(1+\delta^2)x}}~,
\end{equation}
\begin{equation}
G(2;E_0)=-\frac{1}{2J(1+\delta^2)}
+i\Bigl[\frac{x}{2J(1+\delta^2)}+1 \Bigr]
\frac{1}{\sqrt{-x^2-4J(1+\delta^2)x}}~.
\end{equation}
When $E_0<E_i$, i.e., $x>0$, the energy is in the gap,
\begin{equation}
G(0;E_0)=-\frac{1}{\sqrt{x^2+4J(1+\delta^2)x}}~,
\end{equation}
\begin{equation}
G(2;E_0)=-\frac{1}{2J(1+\delta^2)}
+\Bigl[\frac{x}{2J(1+\delta^2)}+1 \Bigr]
\frac{1}{\sqrt{x^2+4J(1+\delta^2)x}}~.
\end{equation}

\section{Diagonalization of Hamiltonian (\ref{inter})}

Hamiltonian (\ref{inter}) can be written as
\begin{equation}
\tilde{H}_0=\sum_k (\eta^{\dag}_{1k}~~\eta^{\dag}_{2k})
\left( \begin{array}{cc}
\epsilon_{k}  &  t_{\perp}  \\
t_{\perp} &  \epsilon_k \\
\end{array}  \right)
\left( \begin{array}{cc}
\eta_{1k}  \\
\eta_{2k} \\
\end{array}  \right)+
\sum_k (\gamma^{\dag}_{1k}~~\gamma^{\dag}_{2k})
\left( \begin{array}{cc}
\tilde{\epsilon}_{k}  &  -t_{\perp}  \\
-t_{\perp} &  \tilde{\epsilon}_k \\
\end{array}  \right)
\left( \begin{array}{cc}
\gamma_{1k}  \\
\gamma_{2k} \\
\end{array}  \right)~.
\end{equation}

Making the transformations
\begin{equation}
\left( \begin{array}{c}
\eta_{1k}\\
\eta_{2k}\\
\end{array}  \right)=\frac{1}{\sqrt{2}}\left( \begin{array}{cc}
1  &  1  \\
-1 &  1 \\
\end{array}  \right)
\left( \begin{array}{c}
\tilde{\eta}_{1k}  \\
\tilde{\eta}_{2k} \\
\end{array}  \right)
\end{equation}
and
\begin{equation}
\left( \begin{array}{c}
\gamma_{1k}\\
\gamma_{2k}\\
\end{array}  \right)=\frac{1}{\sqrt{2}}\left( \begin{array}{cc}
1  &  -1  \\
1 &  1 \\
\end{array}  \right)
\left( \begin{array}{c}
\tilde{\gamma}_{1k}  \\
\tilde{\gamma}_{2k} \\
\end{array}  \right),
\end{equation}
we have
\begin{equation}
\tilde{H}_0=\sum_{Ik}(E^I_{k}\tilde{\eta}^{\dag}_{Ik}\tilde{\eta}_{Ik}
+\tilde{E}^I_k\tilde{\gamma}^{\dag}_{Ik}\tilde{\gamma}_{Ik})
\end{equation}
and the relation between the two types of local operators $\eta_{il}$ 
($\gamma_{il}$) 
and $\tilde{\eta}_{Il}$ ($\tilde{\gamma}_{Il}$):
\begin{eqnarray}
\eta_{1l}&=&\frac{1}{\sqrt{2}}(\tilde{\eta}_{1l}+\tilde{\eta}_{2l})~,\\
\eta_{2l}&=&\frac{1}{\sqrt{2}}(-\tilde{\eta}_{1l}+\tilde{\eta}_{2l})~,\\
\gamma_{1l}&=&\frac{1}{\sqrt{2}}(\tilde{\gamma}_{1l}-\tilde{\gamma}_{2l})~,\\
\gamma_{2l}&=&\frac{1}{\sqrt{2}}(\tilde{\gamma}_{1l}+\tilde{\gamma}_{2l})~.
\end{eqnarray}

\begin{figure}
\caption{Wave functions of the $A_g$ and $B_u$ bound exciton with
$U=10t$, $V=0.5t$, and $\delta=0.2$. The
dashed and solid lines correspond to the $A_g$ and $B_u$
states, respectively.}
\end{figure}

\begin{figure}
\caption{Energy of excitons as a function of 
center-of-mass momentum $K$ with $U=5t$, $V=2t$, and $\delta=0.2$.}
\end{figure}

\begin{figure}
\caption{Optical absorption spectrum with $U=5t$ and $V=t$. 
The $B_u$ exciton occupies 0.520 of the total oscillator strength.
The arrow indicates
the energy of the corresponding $A_g$ exciton.} 
\end{figure}

\begin{figure}
\caption{Optical absorption spectrum with $U=5t$ and $V=2t$. The arrow 
indicates
the position of the $A_g$ exciton, and the $B_u$ exciton takes 0.949 of the 
total transition strength.} 
\end{figure}

\begin{figure}
\caption{Electron-hole correlation function as a function of the impurity 
strength
$V_0$ for impurity potential $H^{\rm imp}_1$ with $U=10t$ and $V=t$. The 
impurity is situated at site 5
in a $N=10$ site chain. The solid, long dashed, and short dashed lines 
correspond to ${\cal R}(5,6)$, ${\cal R}(7,8)$, and ${\cal R}(3,6)$, 
respectively.}
\end{figure}

\begin{figure}
\caption{The hole density at the impurity site vs impurity strength $V_0$ for
the impurity potential $H^{\rm imp}_1$. The parameters here are the same as 
in Fig. 5.}
\end{figure}

\begin{figure}
\caption{Distribution of charge density in real space for the first type of 
impurity, $H^{\rm imp}_1$.
The solid circles describe the hole density, while the open ones are for the 
electron density. Here $U=10t$, $V=t$, and $V_0=0.5t$.}
\end{figure}

\begin{figure}
\caption{Electron-hole correlation function, as a function of the impurity 
strength
$V_0$ for the impurity potential $H^{\rm imp}_2$ with $U=10t$ and $V=t$.
The impurity is located at 
site 5 in a $N=10$ chain. The solid, dashed, and dot-dashed lines illustrate
${\cal R}(5,6)$, ${\cal R}(3,4)$, and ${\cal R}(3,6)$, respectively.}
\end{figure}

\begin{figure}
\caption{Electron-hole correlation function as a function of the impurity 
strength
$V_0$ for the third kind of impurity $H^{\rm imp}_3$ in a $N=10$ chain.
Parameters are the same as for Fig. 8, $U=10t$ and $V=t$.
The solid, dashed, and dot-dashed lines are for
${\cal R}(5,6)$, ${\cal R}(3,4)$, and ${\cal R}(3,6)$, respectively.}
\end{figure}

\begin{figure}
\caption{Distribution $|Z_K|^2$ for the impurity potential $H^{\rm imp}_1$ 
with $U=10t$ and $V=t$ in a
$N=12$ chain. There are two exciton branches ($A_g$ and $B_u$). The solid
symbols indicate the amplitude of excitons in the $A_g$ branch, and the open 
ones
give the amplitude of excitons in the $B_u$ branch. The circle corresponds to
$V_0=0.1t$, the situation before the crossover, and the box corresponds to
$V_0=t$, which is larger than the crossover value.}
\end{figure}

\begin{figure}
\caption{Distribution $|Z_K|^2$ for the second kind of impurity
potential $H^{\rm imp}_2$ with $U=10t$ and $V=t$ in a $N=12$ 
chain. The solid and open symbols have the same meanings as in Fig. 10. The 
circle describes
the case of $V_0=0.1t$ and the box describes $V_0=t$. The line is for the case
$V_0=2t$. The vanishing amplitude in every exciton state shows the breakdown
of the exciton.}
\end{figure}

\begin{figure}
\caption{Distribution $|Z_K|^2$ for the third kind of impurity $H^{\rm imp}_3$
with $U=10t$ and $V=t$ in a $N=12$ chain. The circle and box correspond
to $V_0=0.1t$ and $V_0=t$, respectively.}
\end{figure}

\begin{figure}
\caption{Electron-hole correlation function as a function of the impurity 
strength
$V_0$ for a charged impurity $H^{\rm imp}_4$ in a $N=10$ chain.
Parameters are the same as for Fig. 8, $U=10t$ and $V=t$.
The solid and dashed lines are for
${\cal R}(5,6)$ and ${\cal R}(3,6)$, respectively.}
\end{figure}

\begin{figure}
\caption{Distribution of charge density in real space for a charged 
impurity with extended potential $H^{\rm imp}_4$.
The solid symbols describe the hole density and the open ones are for the 
electron density. The circles and triangles correspond to $V_0=0.5t$ and 
$V_0=5t$, respectively. Here $U=10t$, $V=t$.}
\end{figure}
\begin{figure}

\caption{Distribution $|Z_K|^2$ for a charged impurity with extended
potential $H^{\rm imp}_4$ for $U=10t$ and $V=t$ in a $N=12$ 
chain. The solid and open symbols have the same meaning as in Fig. 10. The 
circle describes the $V_0=0.1t$ case, the box describes $V_0=t$, and the line is 
for $V_0=5t$. The vanishing amplitude in every exciton state indicates the 
dissociation of the exciton.}
\end{figure}

\begin{figure}
\caption{Intrachain and interchain exciton bands with $U=10t$, $V=t$, 
$t_{\perp}=0.5t$ and $V_{\perp}=t$. The two solid lines describe the 
dispersion of 
the interchain exciton, and the four dashed lines are for intrachain excitons.}
\end{figure}

\begin{figure}
\caption{Intrachain and interchain exciton bands with $U=10t$, $V=t$, 
$t_{\perp}=0.2t$ and
$V_{\perp}=1.5t$. Solid lines are the energy spectra for interchain
excitons. Here the interchain exciton is the lowest excited state.}
\end{figure}

\begin{figure}
\caption{Wavefunctions of the lower static interchain exciton with $U=10t$
$V=t$, $t_{\perp}=0.2t$, and $V_{\perp}=t$. The solid line gives the
relative amplitude that the electron and hole are in the same chain, and the 
dashed
line gives the amplitude that they are in different chains.}
\end{figure}

\begin{figure}
\caption{Wavefunctions of the higher static interchain exciton with
$U=10t$, $V=t$, $t_{\perp}=0.2t$, and $V_{\perp}=t$. Here the intrachain
amplitude is zero.}
\end{figure}

\begin{figure}
\caption{Intra/interchain electron-hole correlation vs $r=V_{\perp}/V$ in a 
two-chain
system. Each chain has $N=12$ sites and the parameters are
$U=10t$, $V=t$, and $t_{\perp}=0.2t$. The intrachain correlation function 
illustrated
here by the dashed line is ${\cal R}(5,6)$, and the interchain correlation 
function
given by the solid line is ${\cal R}(5,17)$, where site $12+i$ indicates site
$i$ in the second chain.}
\end{figure}

\begin{figure}
\caption{Schematic diagram of the charge transfer from a polymer chain to
a dopant molecule. The dopant has an acceptor level which interacts with the
polymer by nearest-neighbor hopping $v$.}
\end{figure}

\begin{figure}
\caption{Probability of charge transfer of the $B_u$ state as a function of 
acceptor 
level $\Delta_e$. The parameters in the polymer are $U=10t$, $V=t$, and 
$\delta=0.2$,
corresponding to the energy of the $B_u$ exciton, $E_0=9.537t$ (indicated by 
the arrow). The dopant concentration 
$c=0.2$. In the solid line, 
$v=0.212t$, and in the dashed line, $v=0.566t$.}
\end{figure}

\begin{figure}
\caption{Probability of charge transfer of the $A_g$ state as a function of 
acceptor 
level $\Delta_e$. The parameters are the same as those in Fig. 22, 
and the corresponding $A_g$ state has energy $E_0= 9.140t$ (indicated by the
arrow). The solid and dashed
lines correspond to $v=0.212t$ and $v=0.566t$, respectively.}
\end{figure}


\begin{references}
\bibitem{note} An exact diagonalization for a short chain ($N=6$) with the 
parameters
$V=t$ and $\delta=0.2$ has been carried out (W. Z. Wang, unpublished). 
We found that splittings
between the lowest singlet and triplet exciton states are: $0.034t$ for 
$U=10t$, $0.016t$ for $U=8t$, and $0.14t$ for $U=6t$.
\end{references}
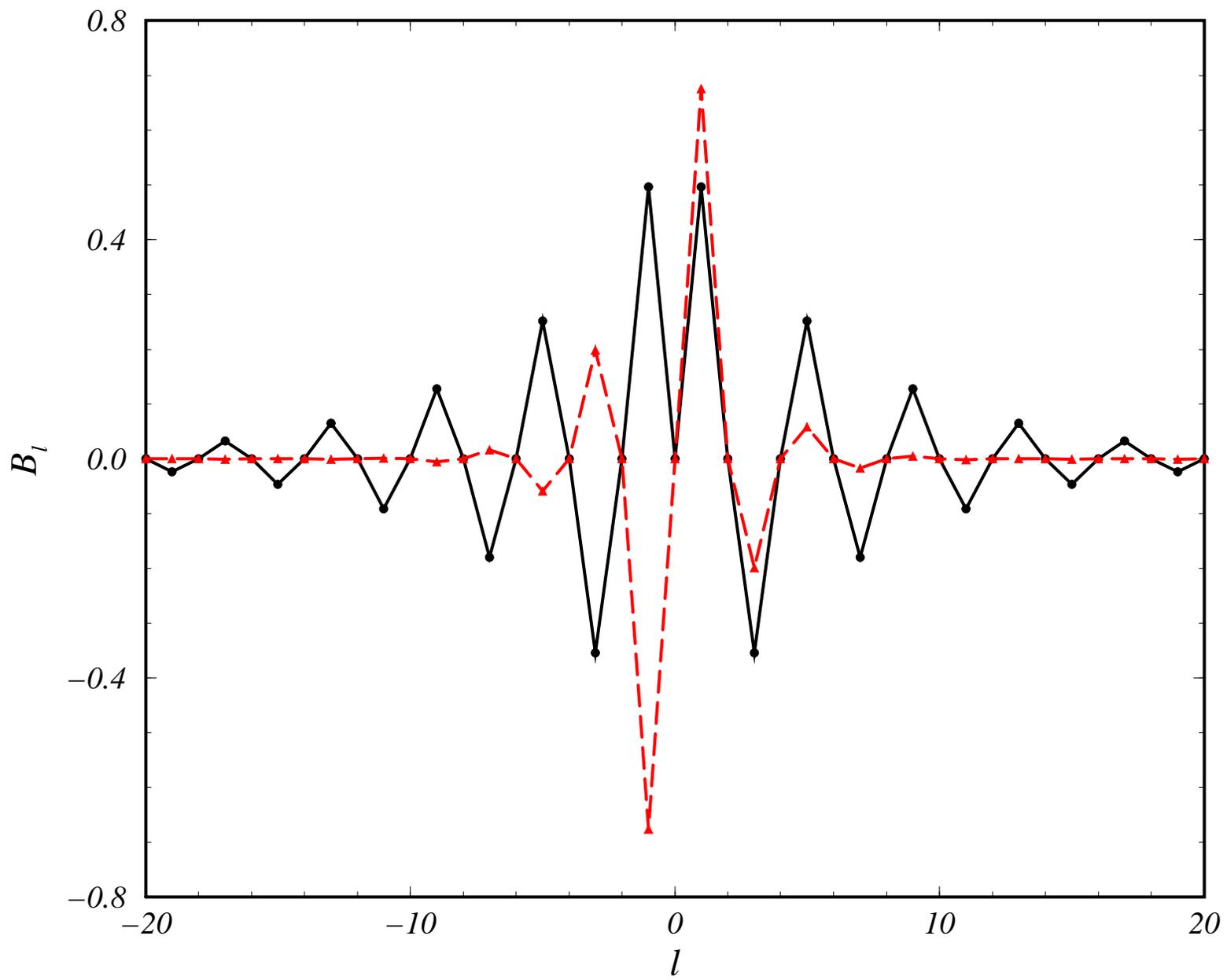
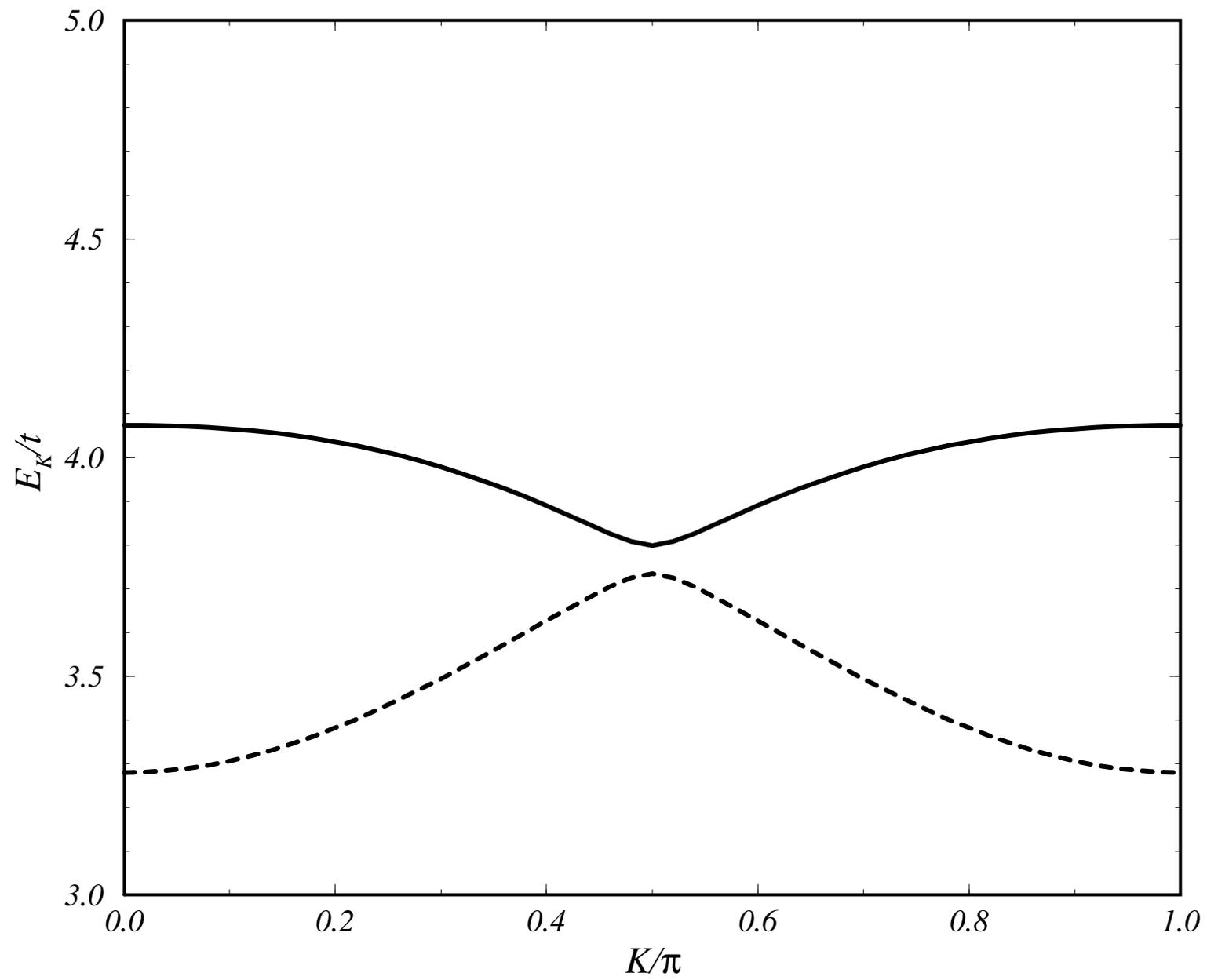
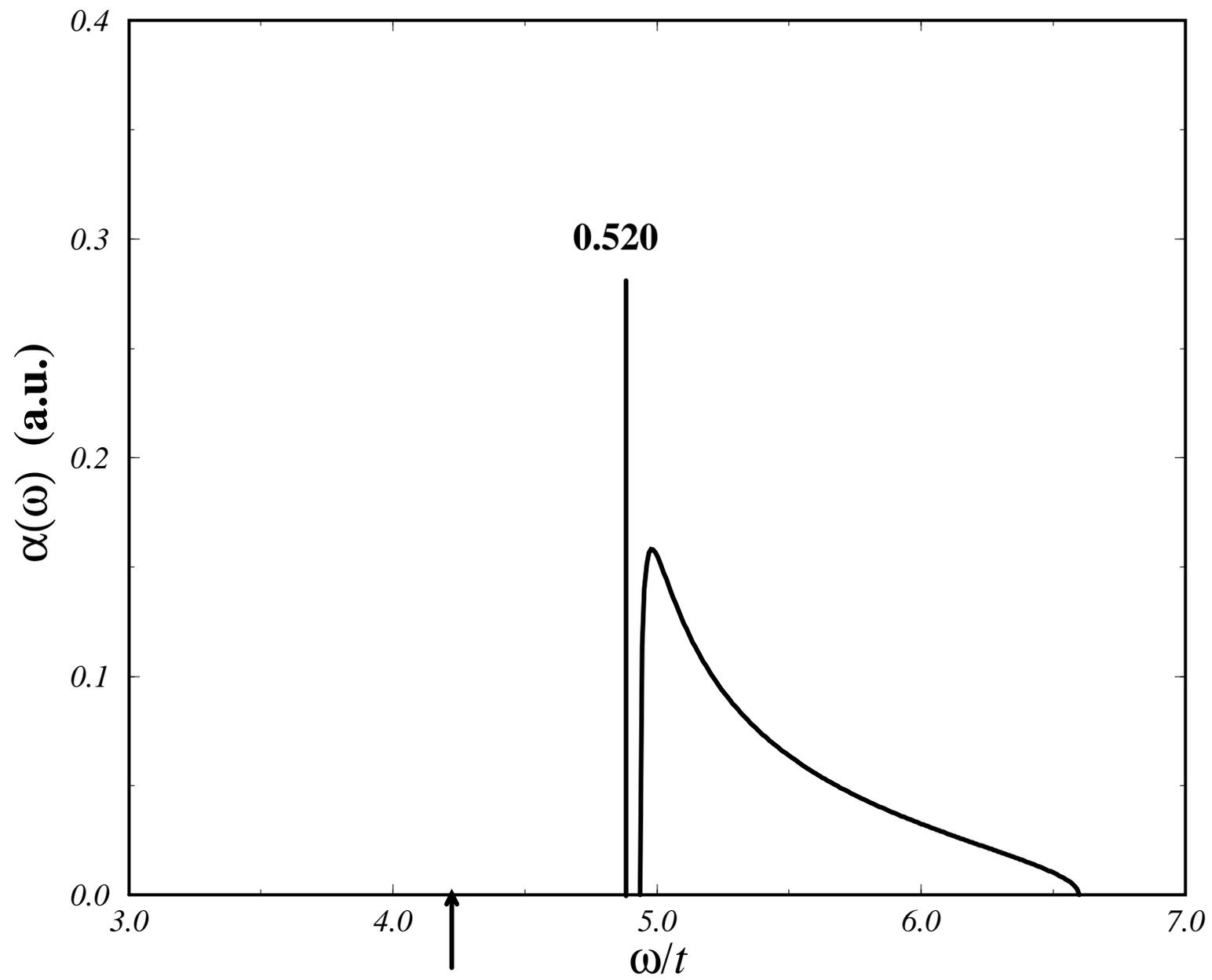
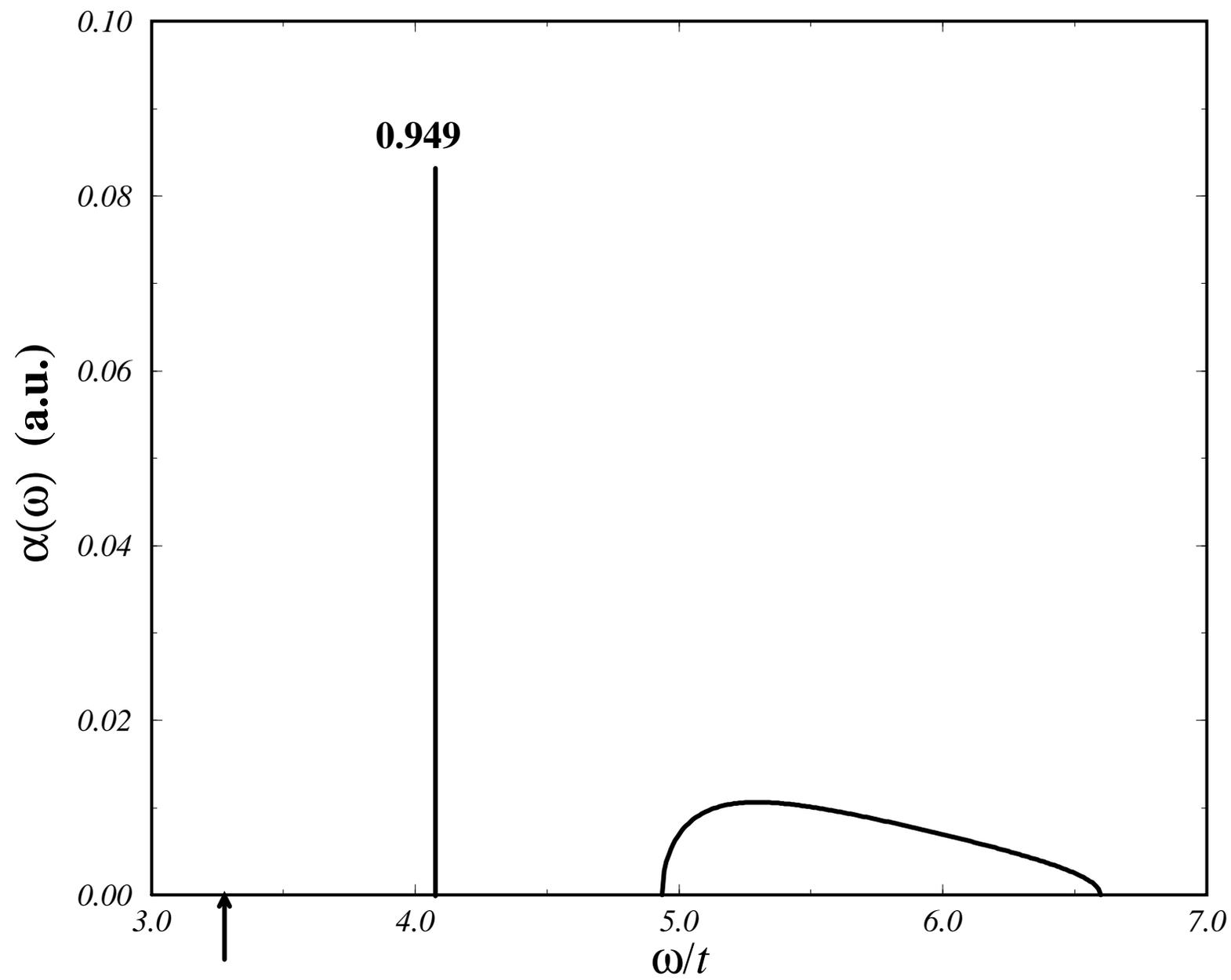
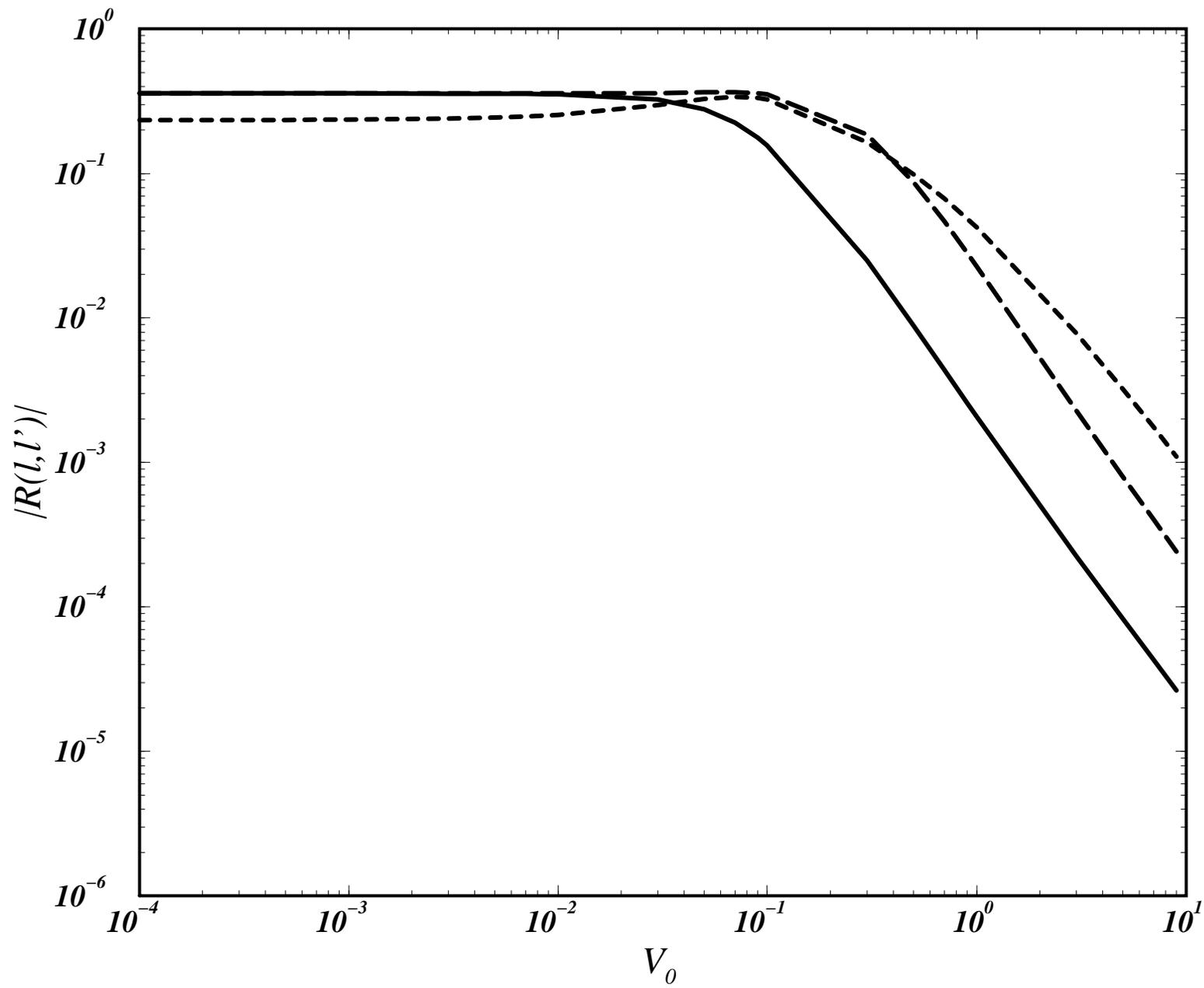
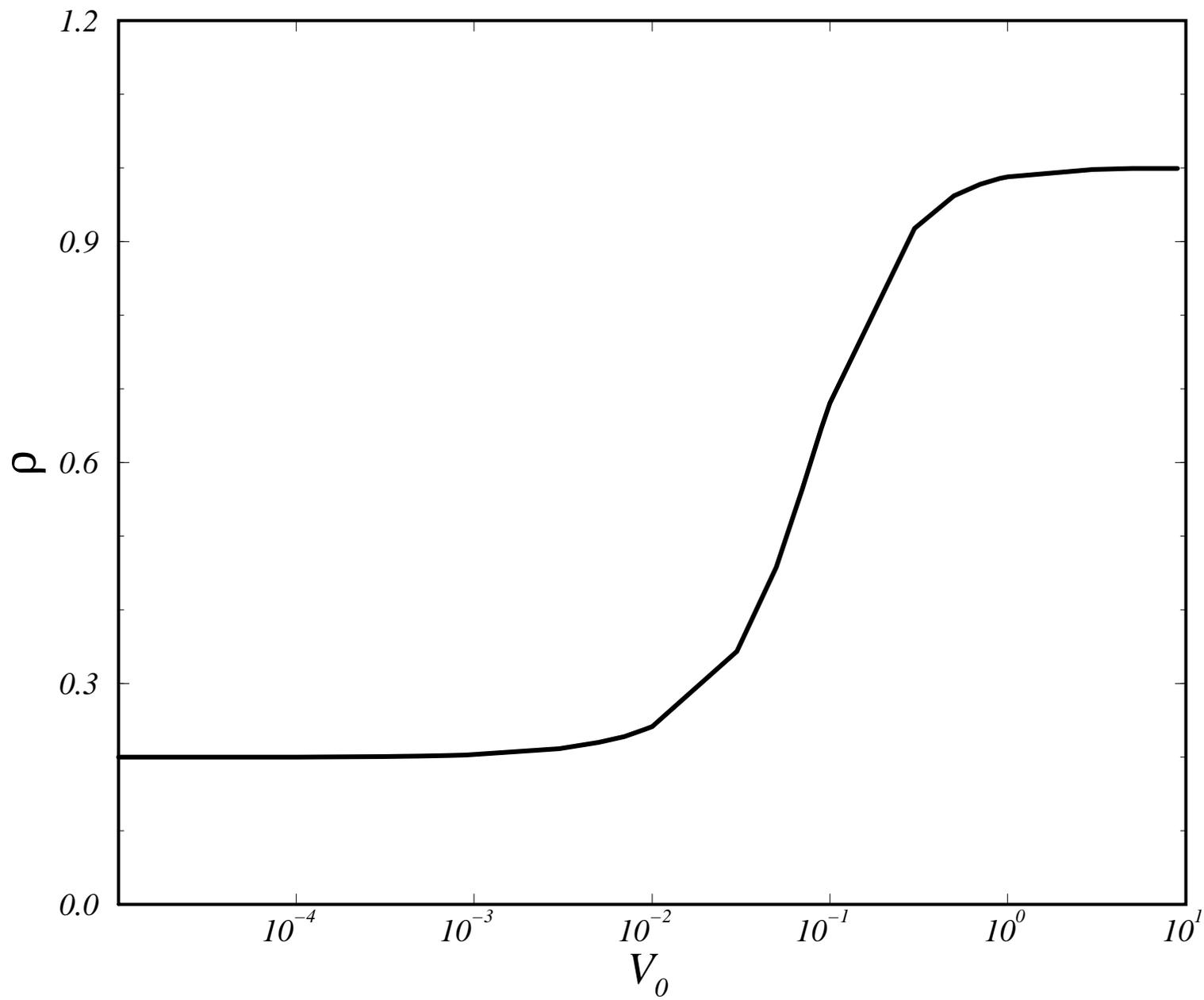
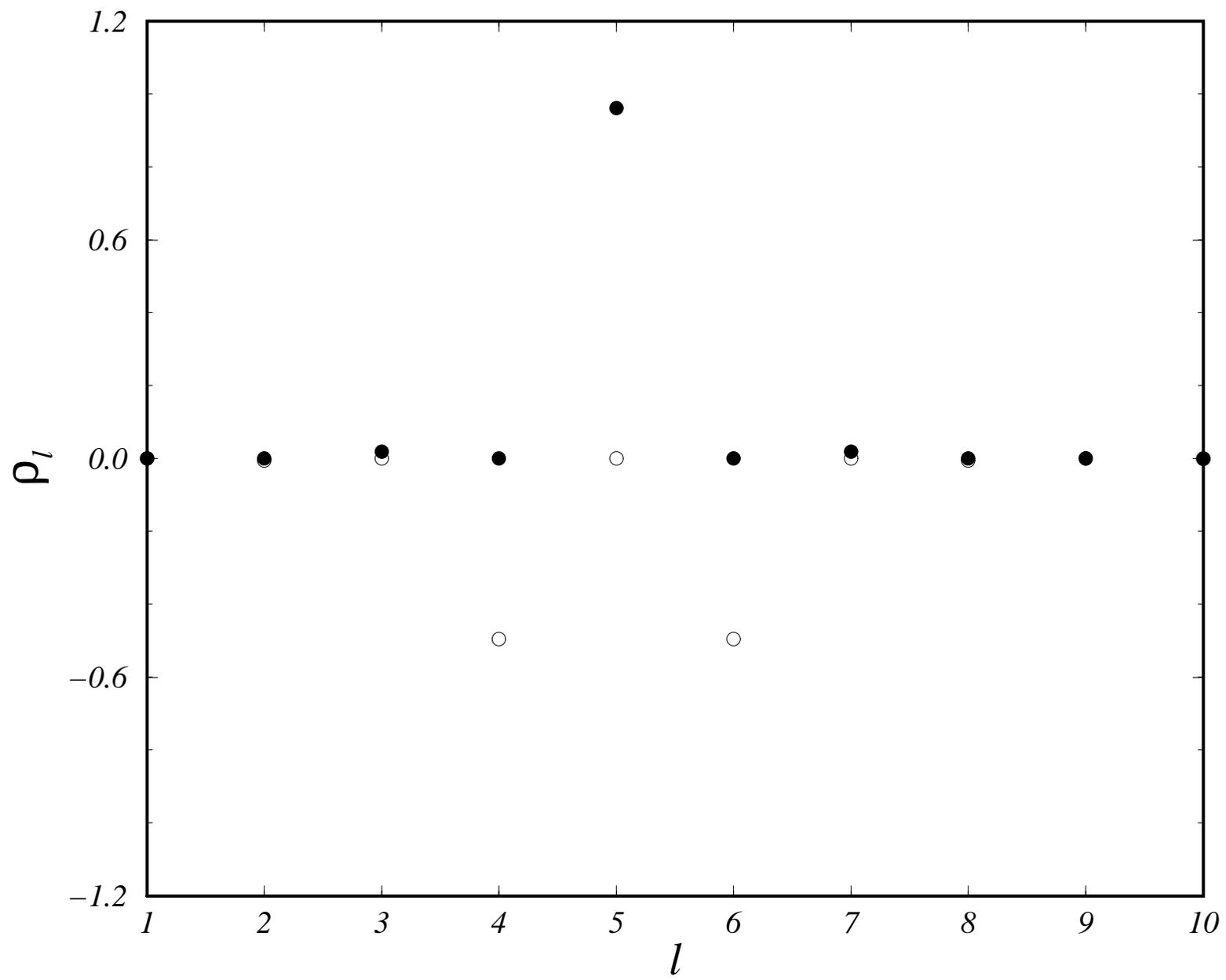
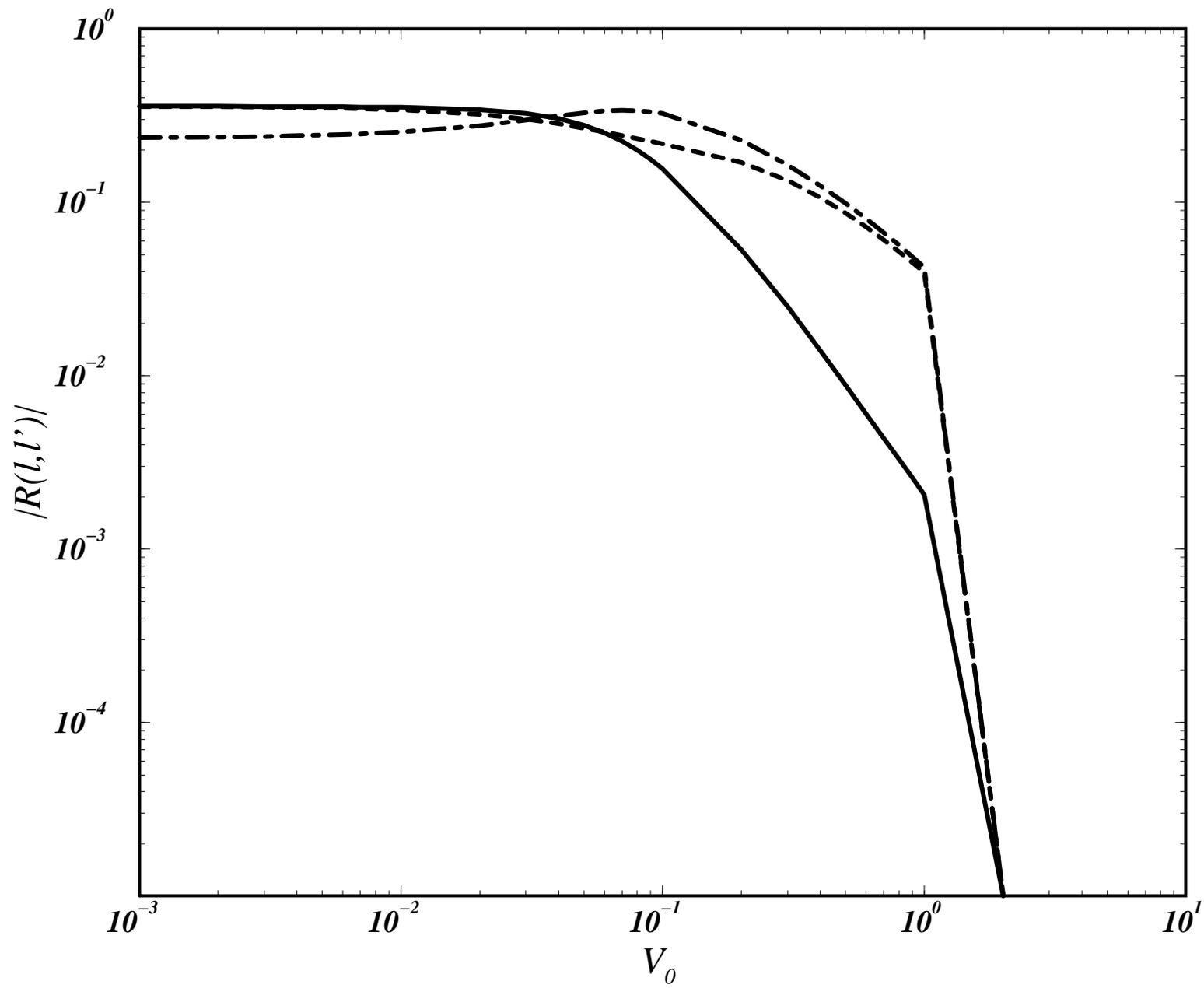
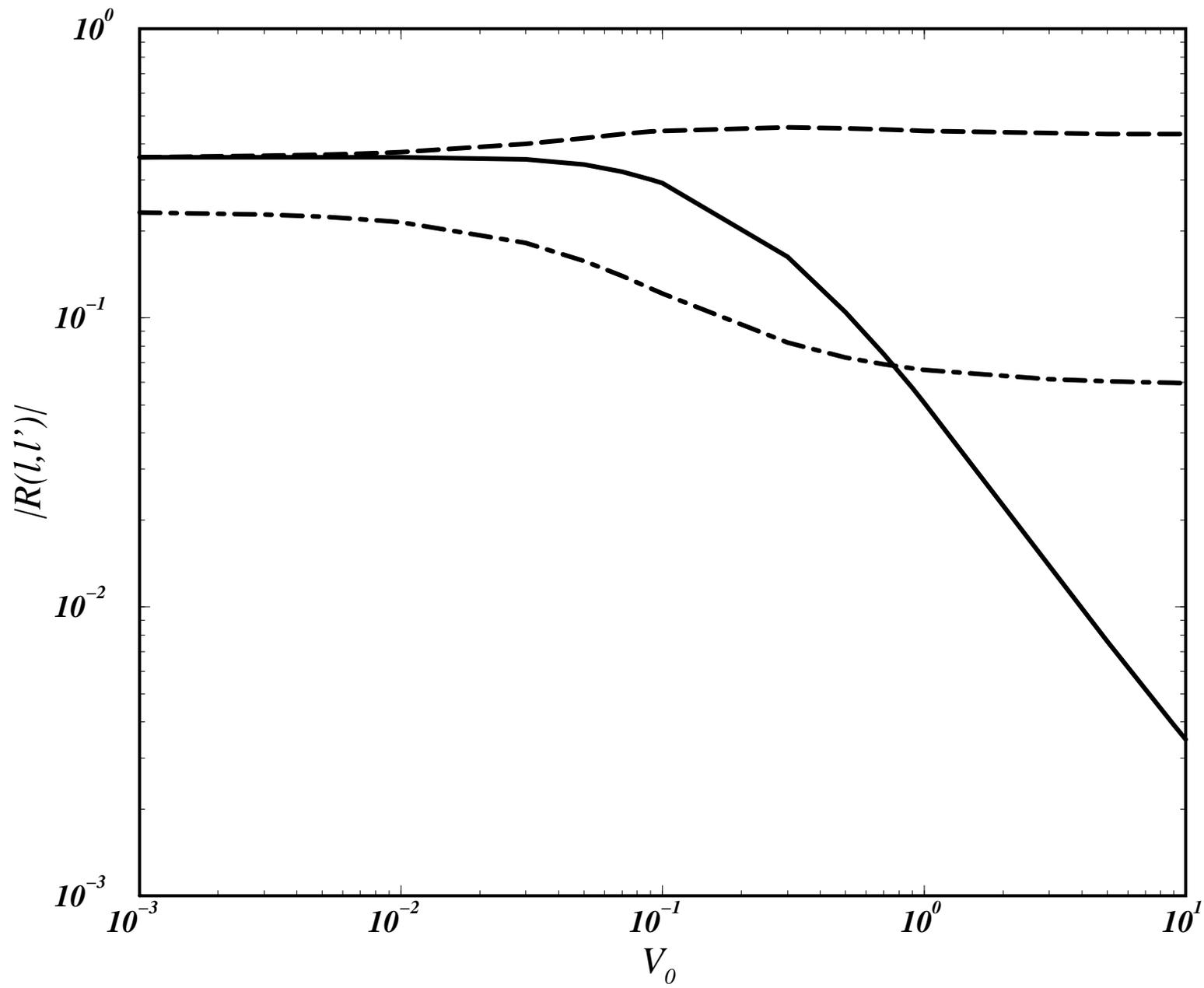
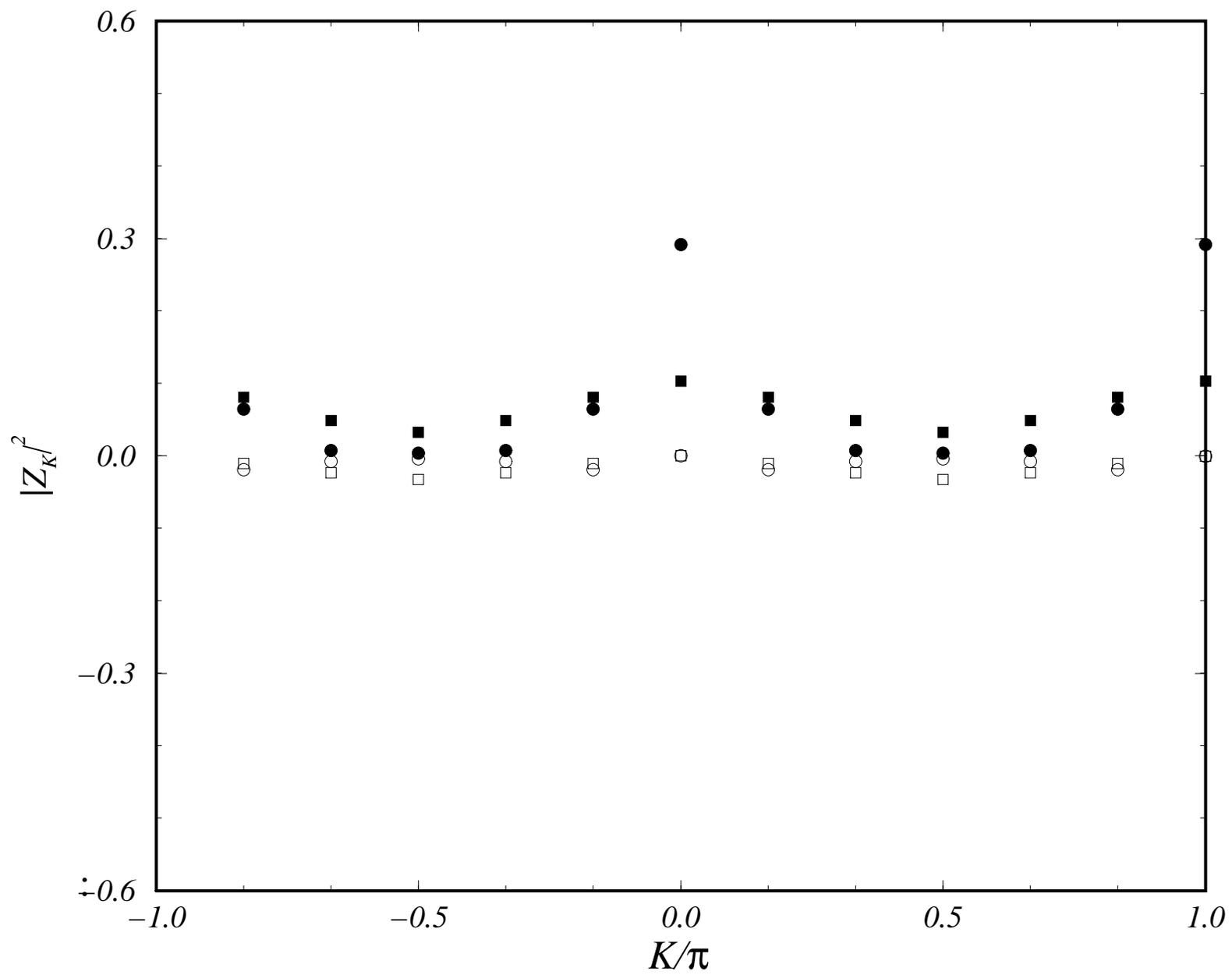
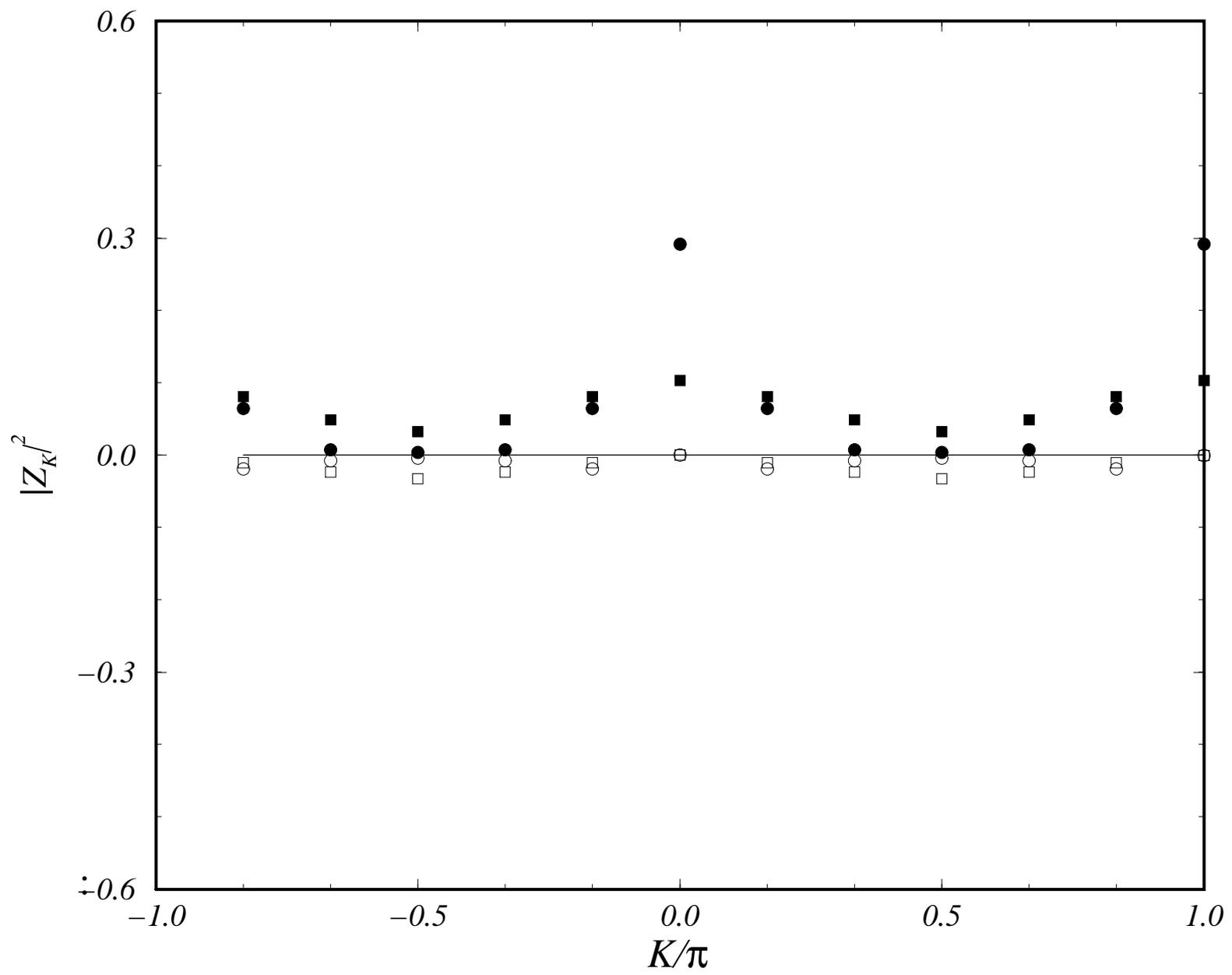
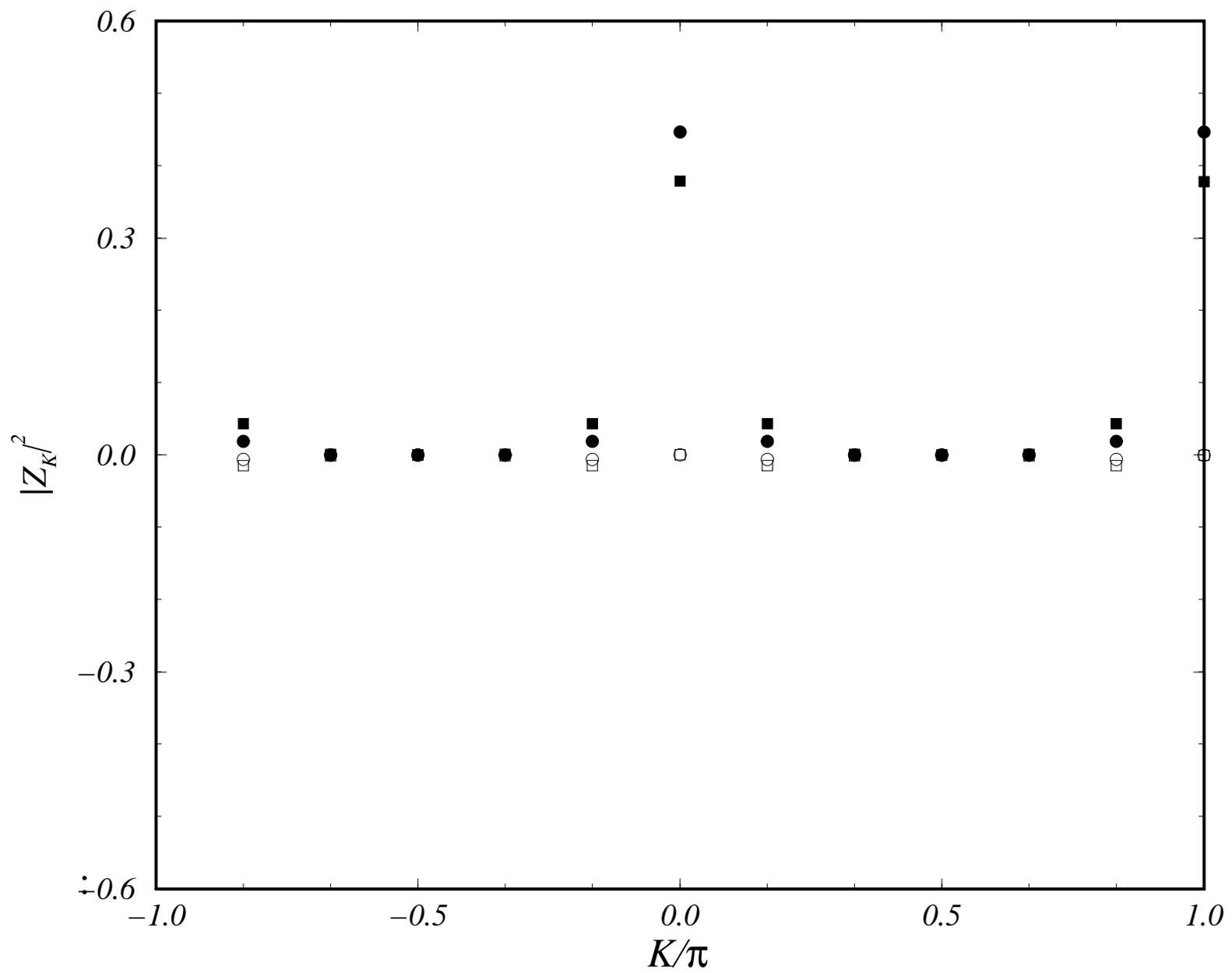
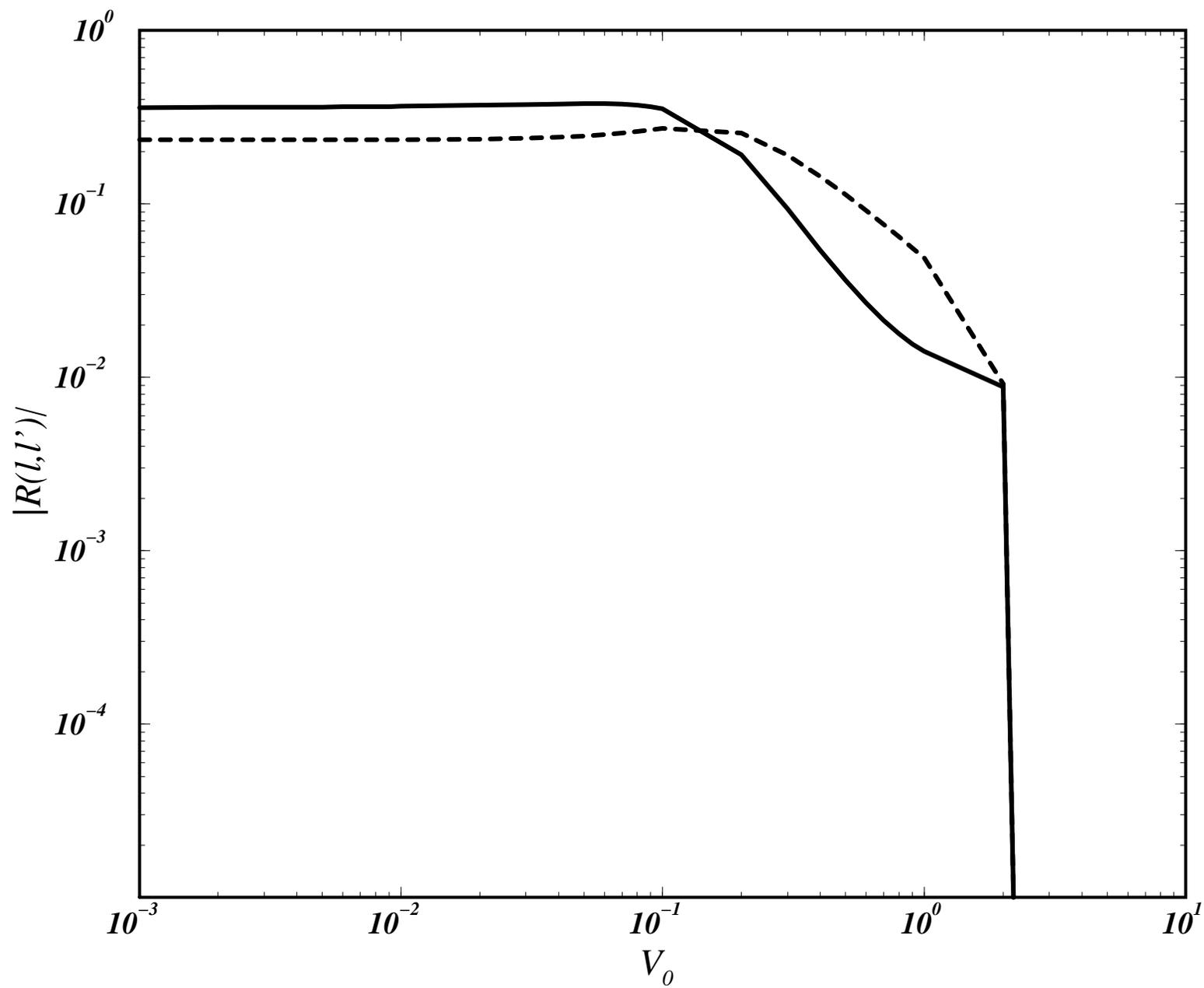
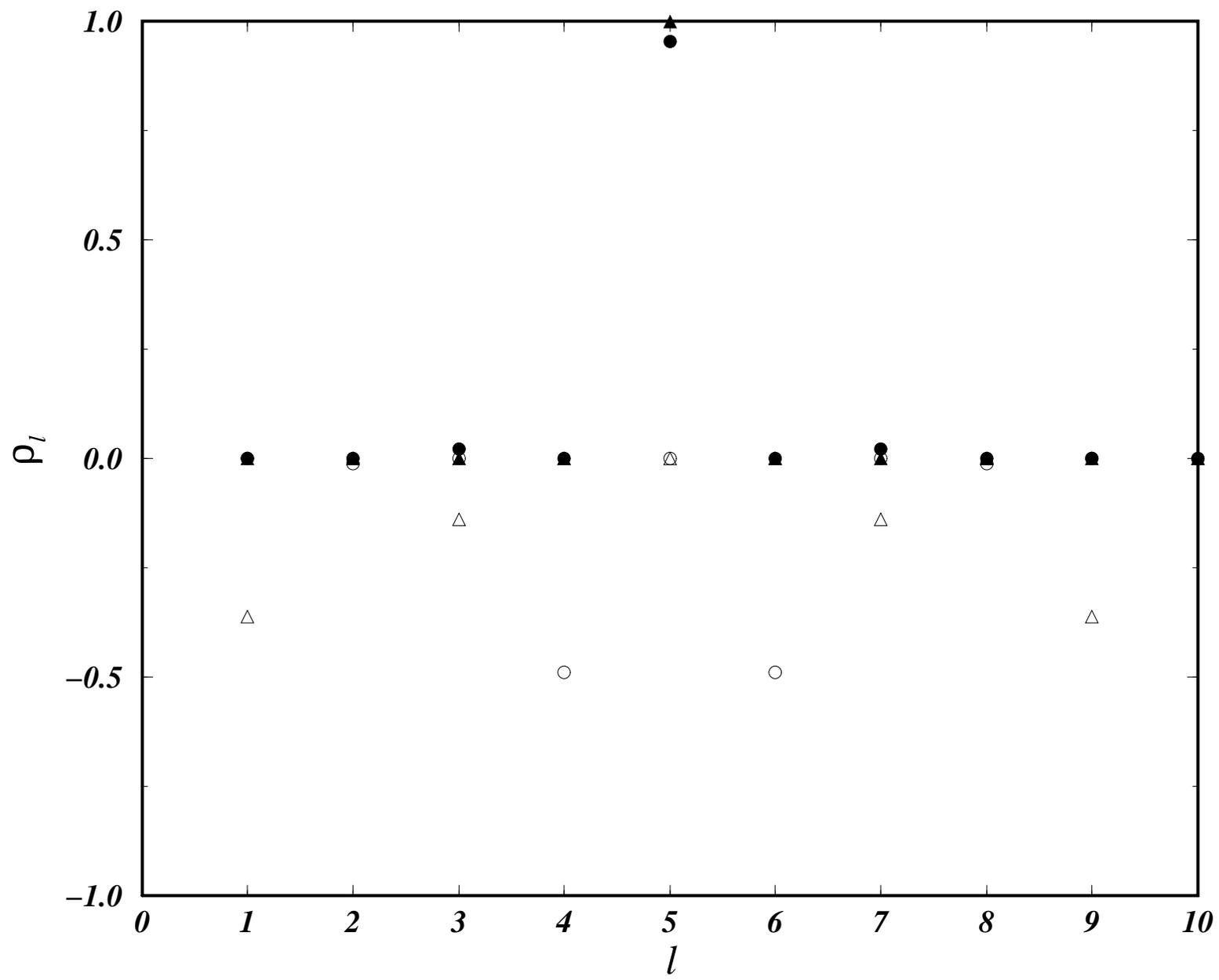
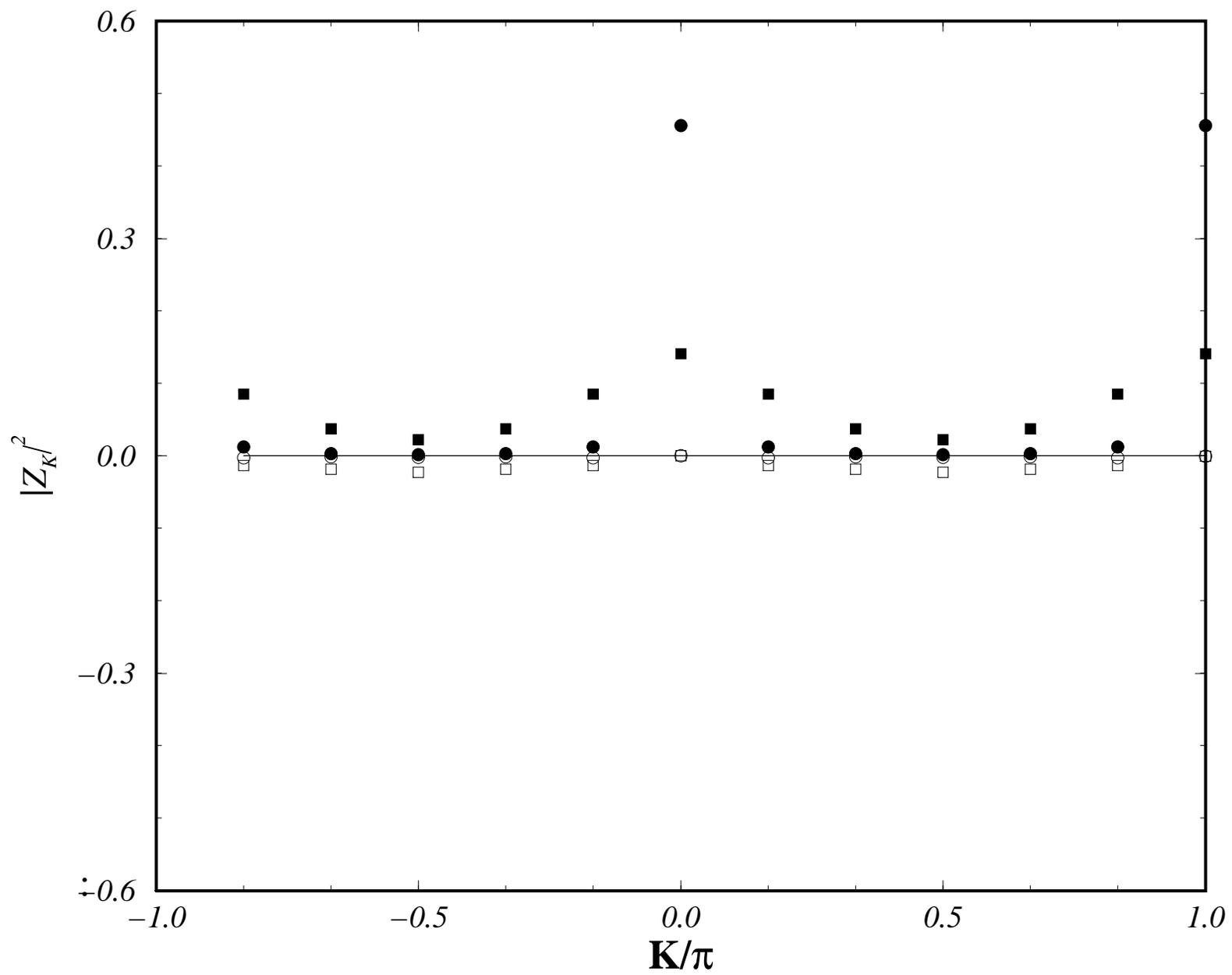
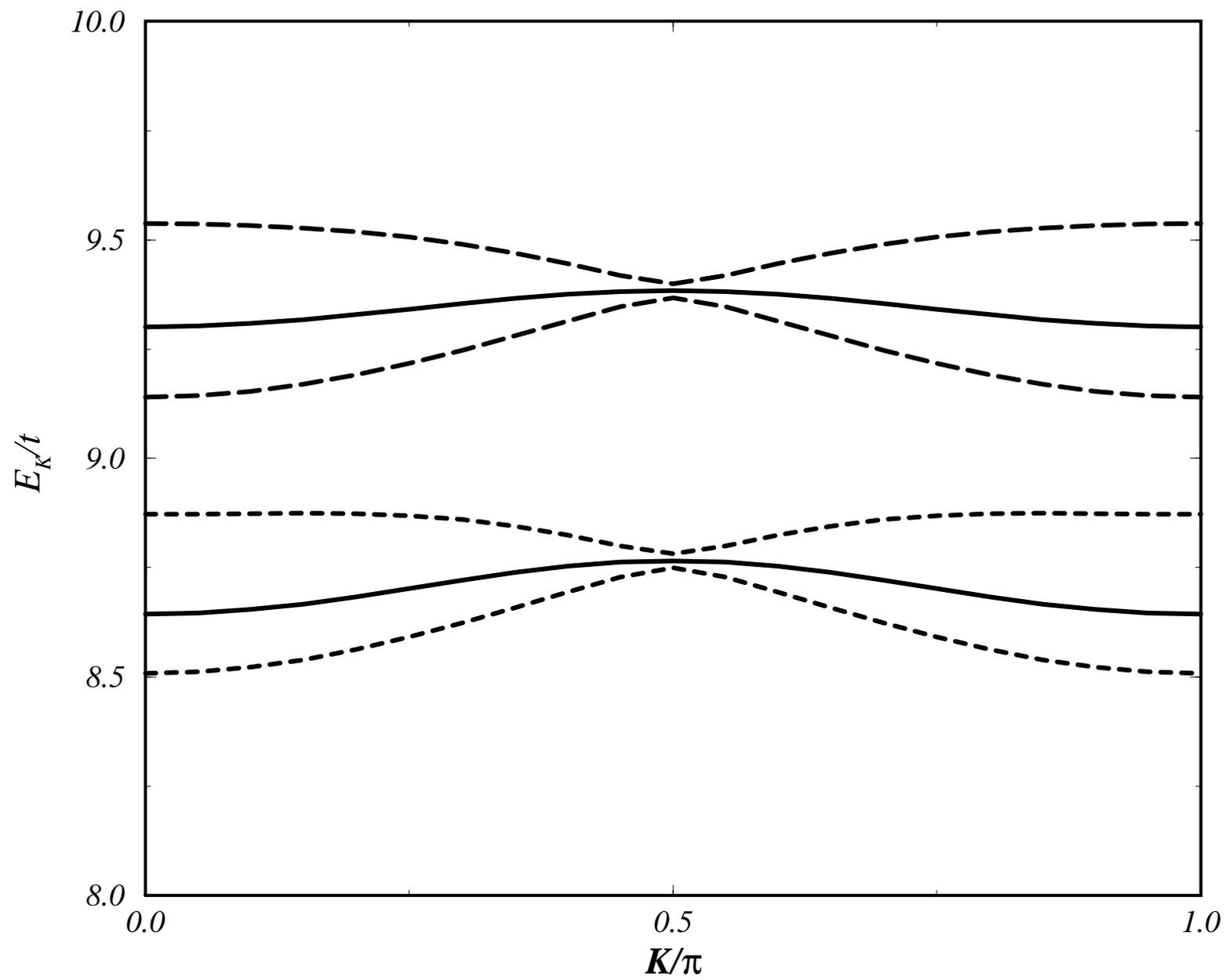
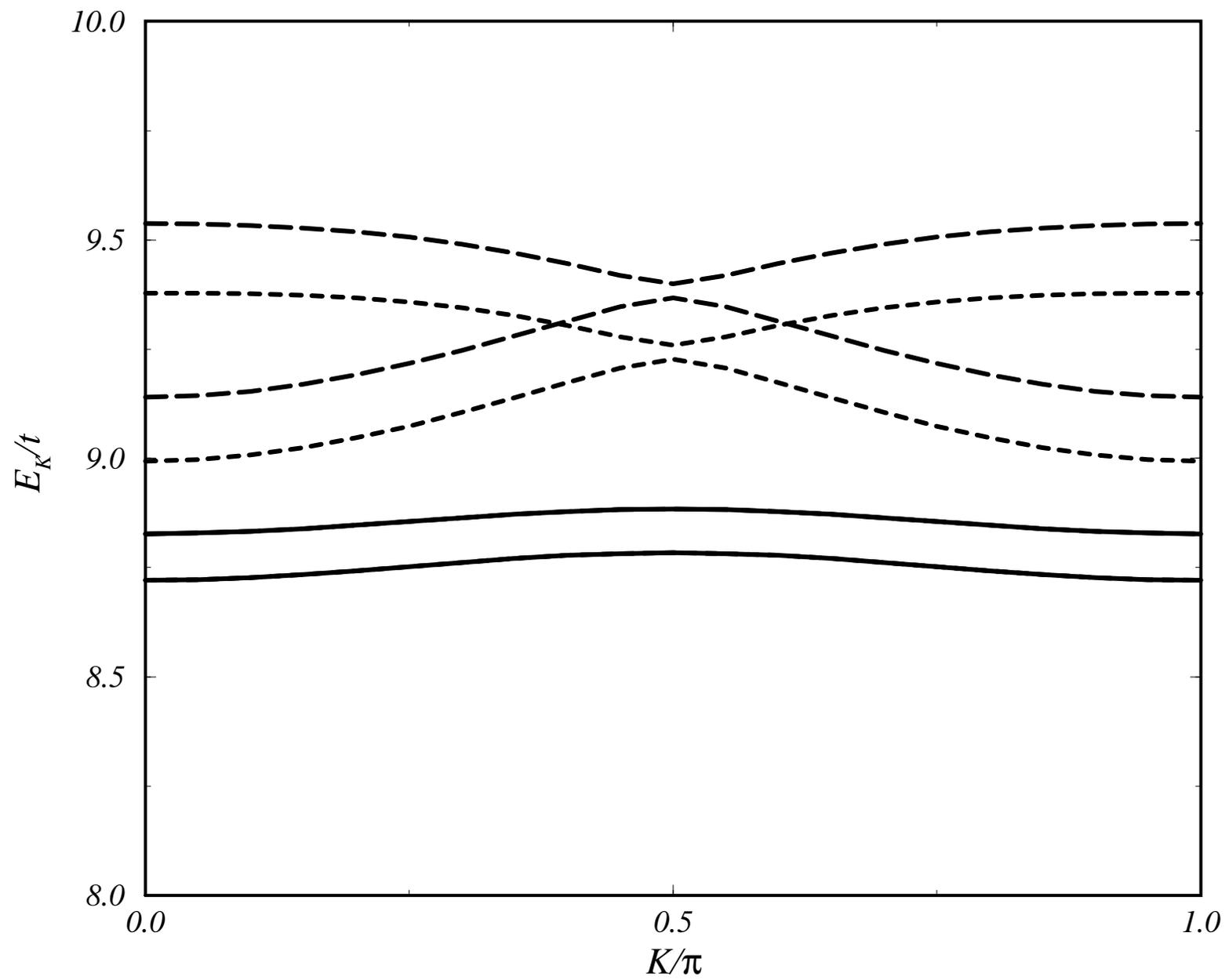
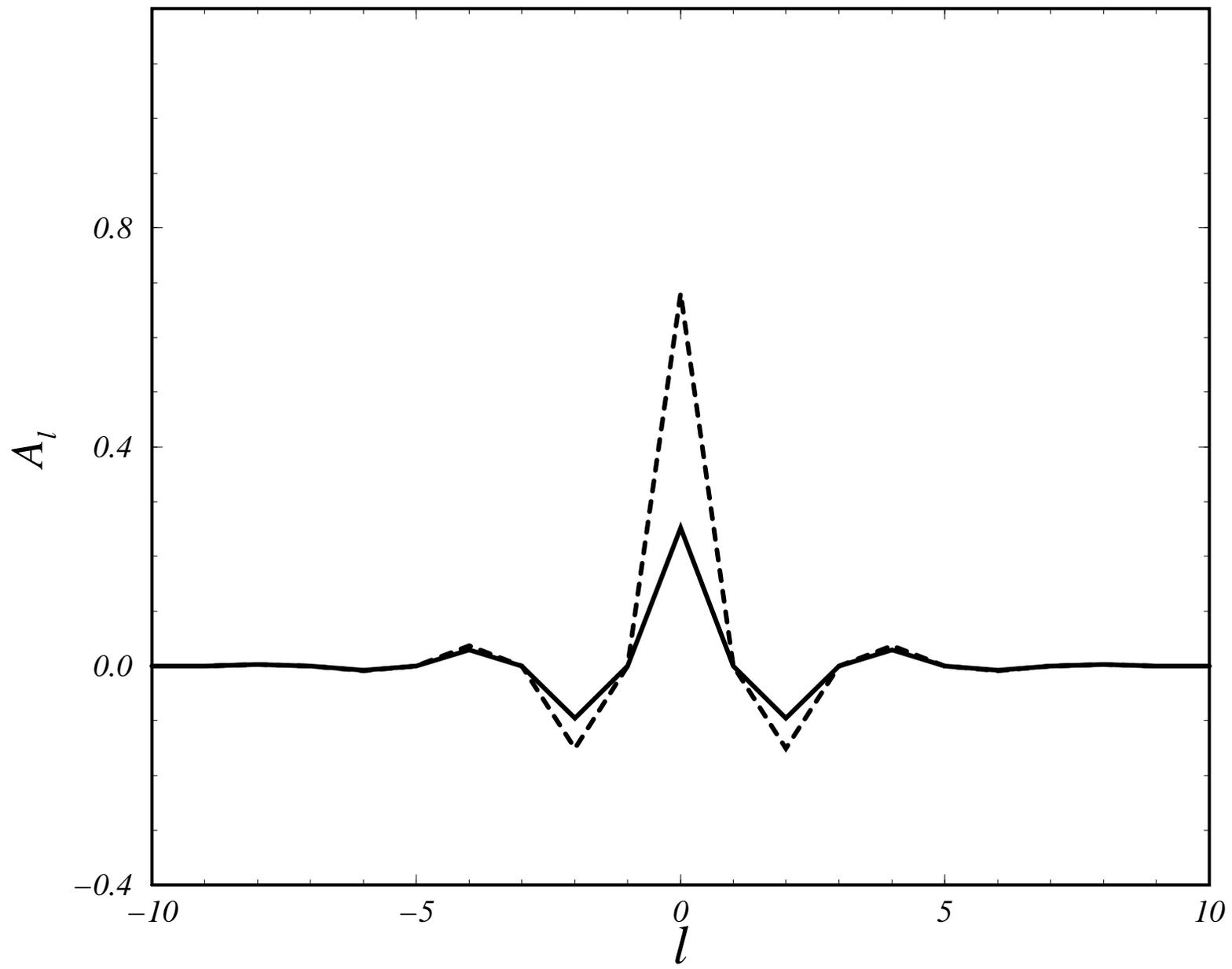
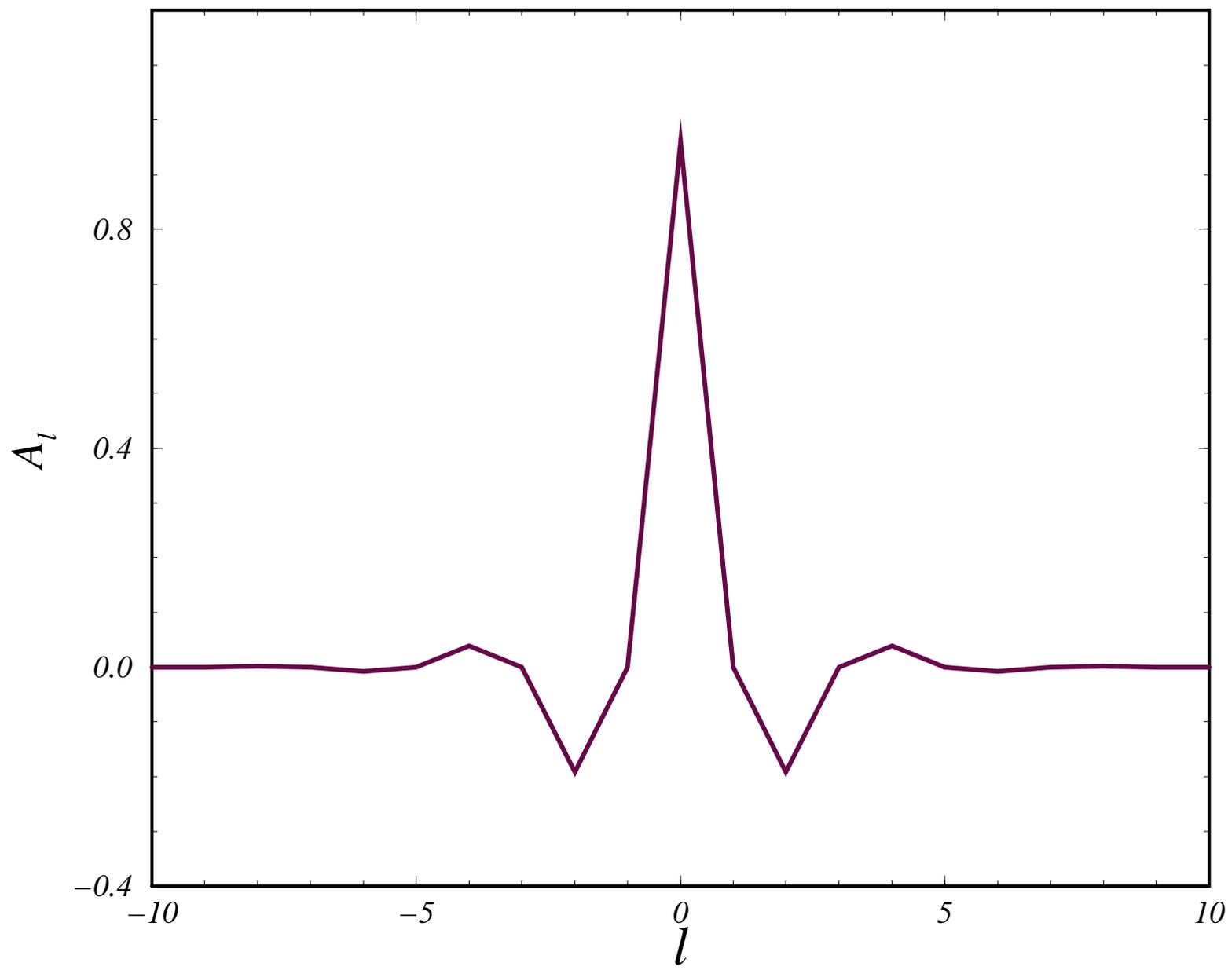
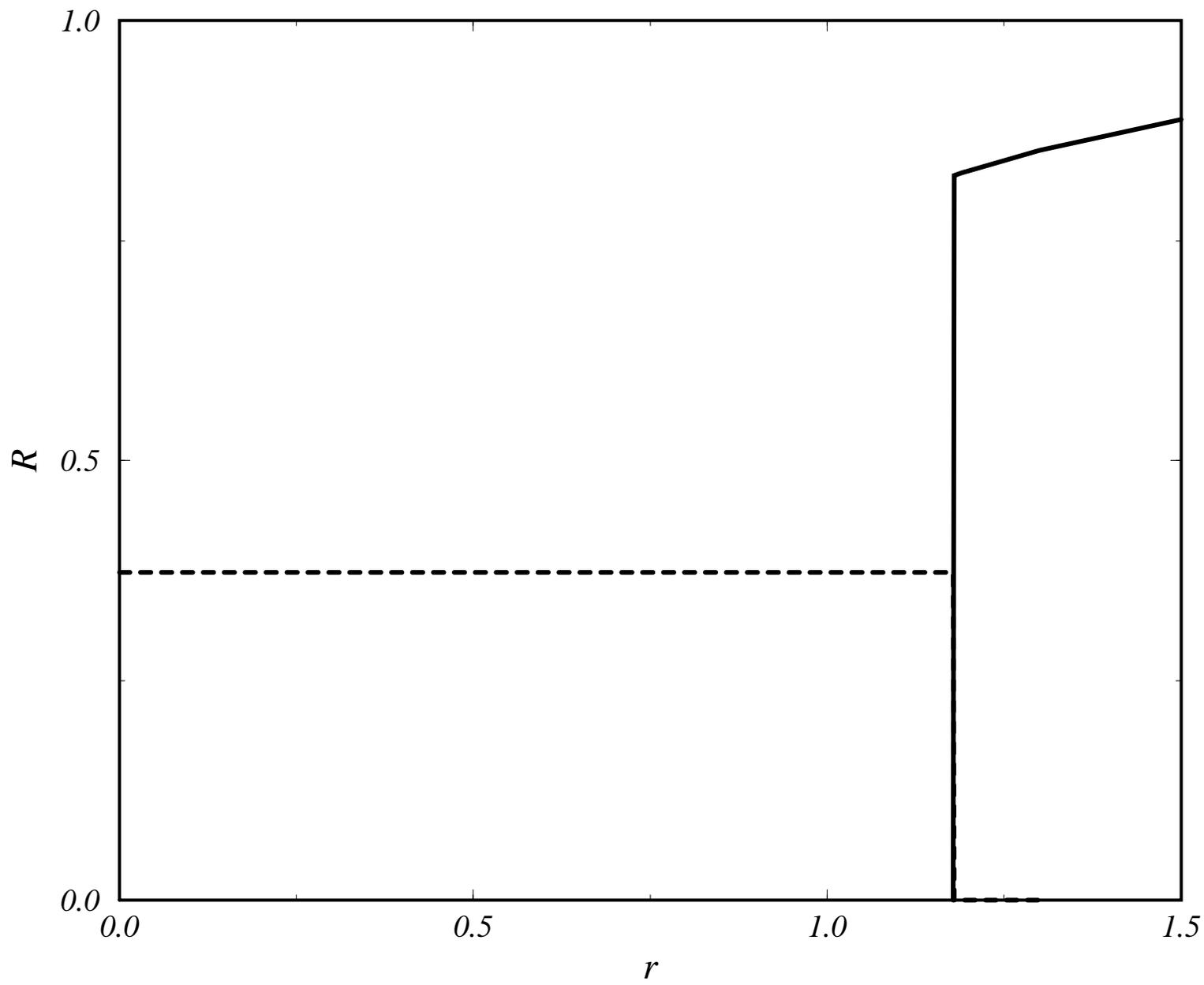
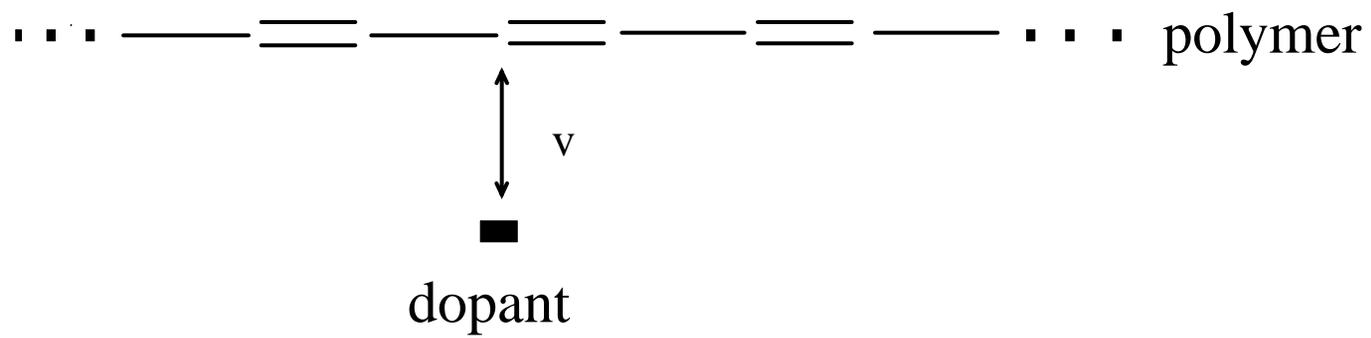
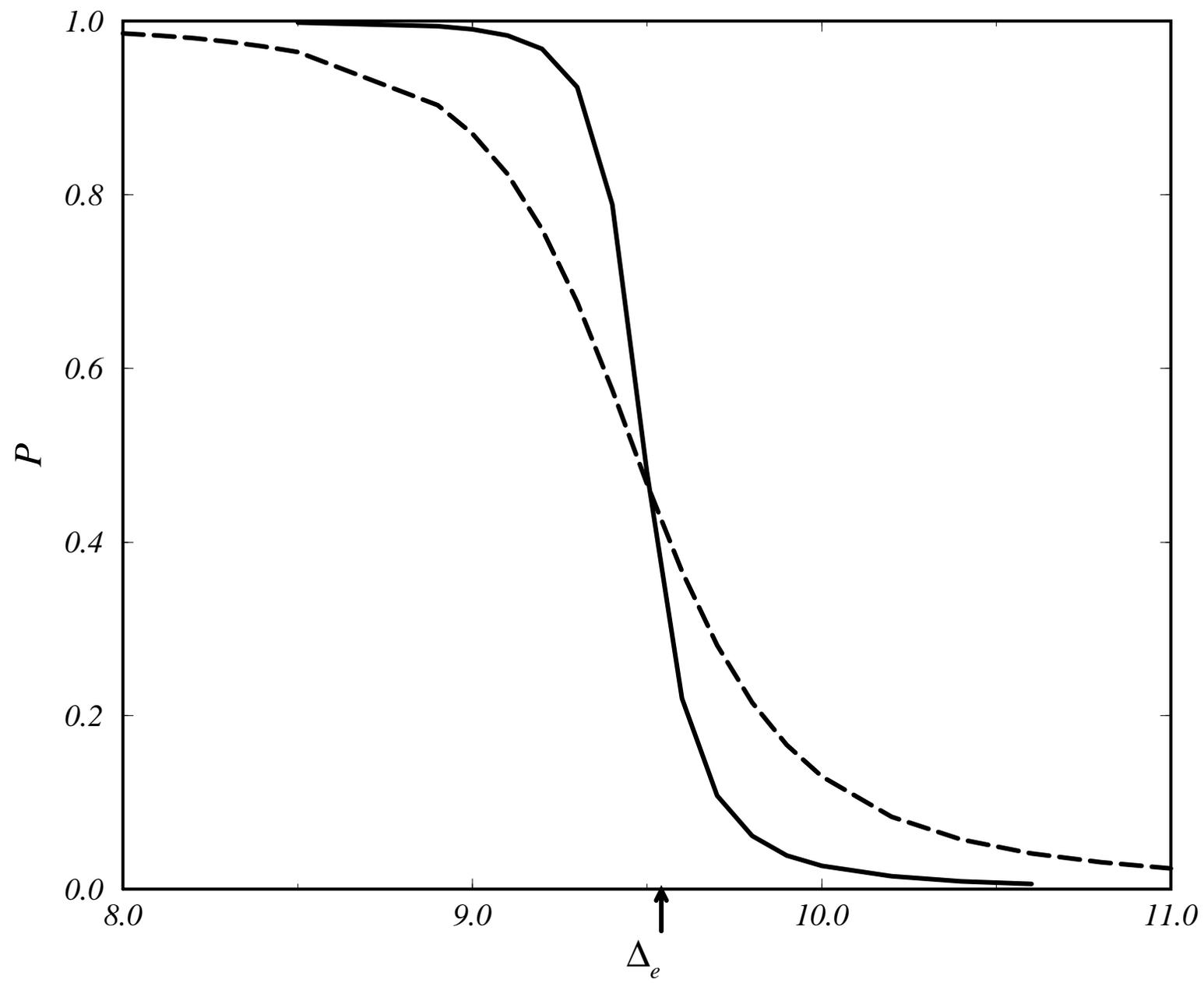
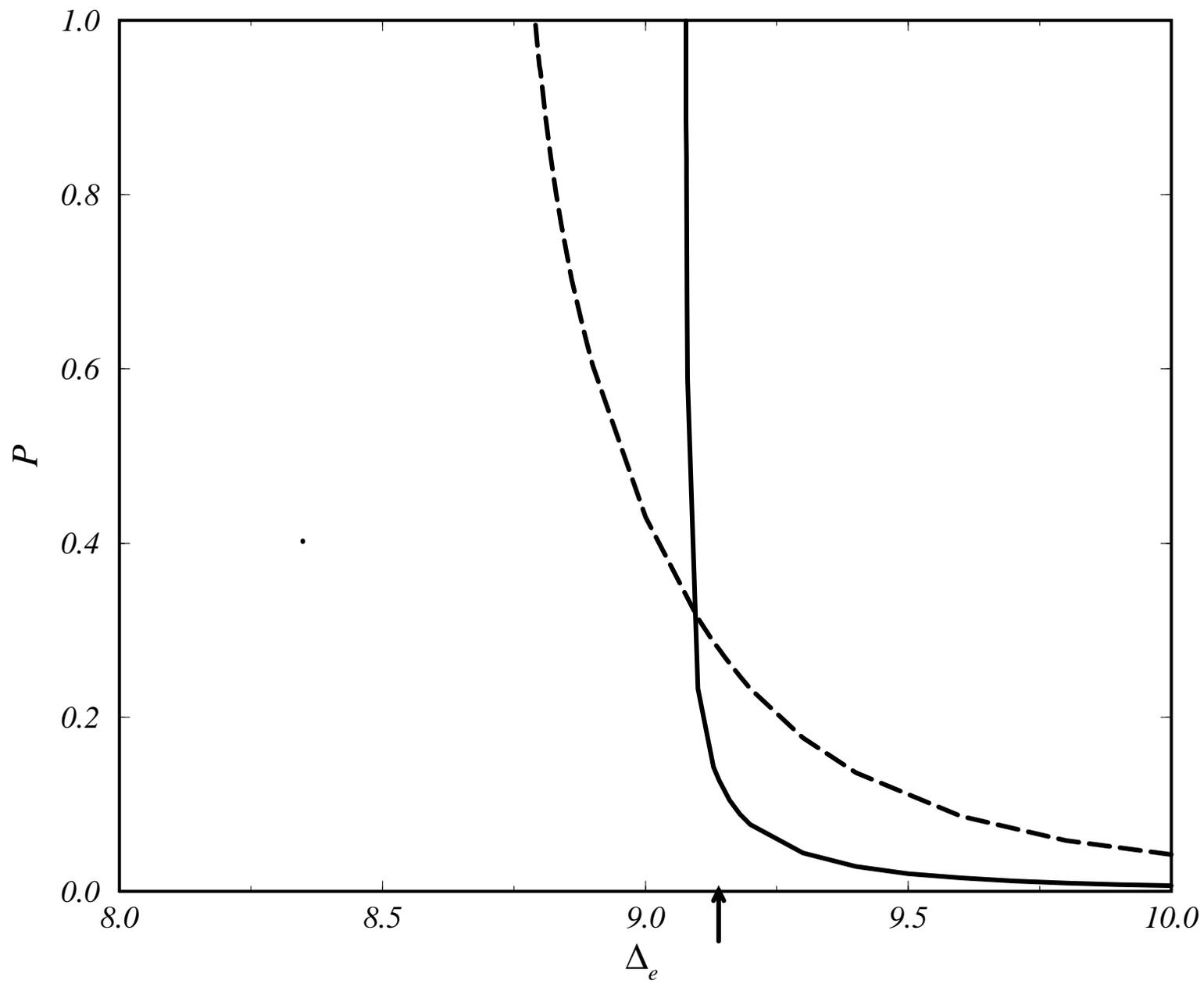
\end{document}